\documentclass[longauth]{aaEC}

\usepackage{graphicx}
\usepackage{natbib}
\usepackage{scalerel}
\usepackage{makecell}

\usepackage[table]{xcolor}

\usepackage{txfonts}
\usepackage[pdfencoding=auto,psdextra]{hyperref}
\hypersetup{
    colorlinks=true,
    linkcolor=blue,
    filecolor=magenta,      
    urlcolor=blue,
    citecolor=blue
}
\makeatletter
\renewcommand*\aa@pageof{, page \thepage{} of \pageref*{LastPage}}
\makeatother

\usepackage[utf8]{inputenc}

\usepackage[switch, modulo]{lineno}
\usepackage{euclid}

\begin{document}
%
% Put the title and authors of your (Standard Project) paper here
%
\title{Euclid Quick Data Release (Q1)}
\subtitle{An investigation of optically faint, red objects in the Euclid Deep Fields}    
%\title{\Euclid\/: HIERO' characterization in the Euclid Deep Fields \thanks{This paper is published on
%       behalf of the Euclid Consortium}}

%% please do not edit the author list once you copy it from the
%% Publication Portal -- contact ECEB Bureau for changes
   
%%%% Version Monday 17th of March 2025 05:25:00 PM UT												
%%%% Please do not edit the author list -- contact ECEB Bureau for changes
\newcommand{\orcid}[1]{} %% if already defined in aa.cls: comment, or use renewcommand			   
\author{Euclid Collaboration: G.~Girardi\orcid{0009-0005-6156-4066}\thanks{\email{giorgia.girardi.1@phd.unipd.it}}\inst{\ref{aff1},\ref{aff2}}
\and G.~Rodighiero\orcid{0000-0002-9415-2296}\inst{\ref{aff1},\ref{aff2}}
\and L.~Bisigello\orcid{0000-0003-0492-4924}\inst{\ref{aff2}}
\and A.~Enia\orcid{0000-0002-0200-2857}\inst{\ref{aff3},\ref{aff4}}
\and A.~Grazian\orcid{0000-0002-5688-0663}\inst{\ref{aff2}}
\and E.~Dalla~Bont\`a\orcid{0000-0001-9931-8681}\inst{\ref{aff1},\ref{aff2},\ref{aff5}}
\and E.~Daddi\orcid{0000-0002-3331-9590}\inst{\ref{aff6}}
\and S.~Serjeant\orcid{0000-0002-0517-7943}\inst{\ref{aff7}}
\and G.~Gandolfi\orcid{0000-0003-3248-5666}\inst{\ref{aff8},\ref{aff2}}
\and C.~C.~Lovell\orcid{0000-0001-7964-5933}\inst{\ref{aff9}}
\and K.~I.~Caputi\orcid{0000-0001-8183-1460}\inst{\ref{aff10},\ref{aff11}}
\and A.~Bianchetti\orcid{0009-0002-8916-3430}\inst{\ref{aff1},\ref{aff2}}
\and A.~Vietri\orcid{0000-0003-4032-0853}\inst{\ref{aff1}}
\and N.~Aghanim\orcid{0000-0002-6688-8992}\inst{\ref{aff12}}
\and B.~Altieri\orcid{0000-0003-3936-0284}\inst{\ref{aff13}}
\and A.~Amara\inst{\ref{aff14}}
\and S.~Andreon\orcid{0000-0002-2041-8784}\inst{\ref{aff15}}
\and N.~Auricchio\orcid{0000-0003-4444-8651}\inst{\ref{aff4}}
\and H.~Aussel\orcid{0000-0002-1371-5705}\inst{\ref{aff6}}
\and C.~Baccigalupi\orcid{0000-0002-8211-1630}\inst{\ref{aff16},\ref{aff17},\ref{aff18},\ref{aff19}}
\and M.~Baldi\orcid{0000-0003-4145-1943}\inst{\ref{aff3},\ref{aff4},\ref{aff20}}
\and A.~Balestra\orcid{0000-0002-6967-261X}\inst{\ref{aff2}}
\and S.~Bardelli\orcid{0000-0002-8900-0298}\inst{\ref{aff4}}
\and P.~Battaglia\orcid{0000-0002-7337-5909}\inst{\ref{aff4}}
\and R.~Bender\orcid{0000-0001-7179-0626}\inst{\ref{aff21},\ref{aff22}}
\and A.~Biviano\orcid{0000-0002-0857-0732}\inst{\ref{aff17},\ref{aff16}}
\and A.~Bonchi\orcid{0000-0002-2667-5482}\inst{\ref{aff23}}
\and E.~Branchini\orcid{0000-0002-0808-6908}\inst{\ref{aff24},\ref{aff25},\ref{aff15}}
\and M.~Brescia\orcid{0000-0001-9506-5680}\inst{\ref{aff26},\ref{aff27}}
\and J.~Brinchmann\orcid{0000-0003-4359-8797}\inst{\ref{aff28},\ref{aff29}}
\and S.~Camera\orcid{0000-0003-3399-3574}\inst{\ref{aff30},\ref{aff31},\ref{aff32}}
\and G.~Ca\~nas-Herrera\orcid{0000-0003-2796-2149}\inst{\ref{aff33},\ref{aff34},\ref{aff35}}
\and V.~Capobianco\orcid{0000-0002-3309-7692}\inst{\ref{aff32}}
\and C.~Carbone\orcid{0000-0003-0125-3563}\inst{\ref{aff36}}
\and J.~Carretero\orcid{0000-0002-3130-0204}\inst{\ref{aff37},\ref{aff38}}
\and S.~Casas\orcid{0000-0002-4751-5138}\inst{\ref{aff39}}
\and M.~Castellano\orcid{0000-0001-9875-8263}\inst{\ref{aff40}}
\and G.~Castignani\orcid{0000-0001-6831-0687}\inst{\ref{aff4}}
\and S.~Cavuoti\orcid{0000-0002-3787-4196}\inst{\ref{aff27},\ref{aff41}}
\and K.~C.~Chambers\orcid{0000-0001-6965-7789}\inst{\ref{aff42}}
\and A.~Cimatti\inst{\ref{aff43}}
\and C.~Colodro-Conde\inst{\ref{aff44}}
\and G.~Congedo\orcid{0000-0003-2508-0046}\inst{\ref{aff45}}
\and C.~J.~Conselice\orcid{0000-0003-1949-7638}\inst{\ref{aff46}}
\and L.~Conversi\orcid{0000-0002-6710-8476}\inst{\ref{aff47},\ref{aff13}}
\and Y.~Copin\orcid{0000-0002-5317-7518}\inst{\ref{aff48}}
\and F.~Courbin\orcid{0000-0003-0758-6510}\inst{\ref{aff49},\ref{aff50}}
\and H.~M.~Courtois\orcid{0000-0003-0509-1776}\inst{\ref{aff51}}
\and M.~Cropper\orcid{0000-0003-4571-9468}\inst{\ref{aff52}}
\and A.~Da~Silva\orcid{0000-0002-6385-1609}\inst{\ref{aff53},\ref{aff54}}
\and H.~Degaudenzi\orcid{0000-0002-5887-6799}\inst{\ref{aff55}}
\and G.~De~Lucia\orcid{0000-0002-6220-9104}\inst{\ref{aff17}}
\and A.~M.~Di~Giorgio\orcid{0000-0002-4767-2360}\inst{\ref{aff56}}
\and C.~Dolding\orcid{0009-0003-7199-6108}\inst{\ref{aff52}}
\and H.~Dole\orcid{0000-0002-9767-3839}\inst{\ref{aff12}}
\and F.~Dubath\orcid{0000-0002-6533-2810}\inst{\ref{aff55}}
\and C.~A.~J.~Duncan\orcid{0009-0003-3573-0791}\inst{\ref{aff46}}
\and X.~Dupac\inst{\ref{aff13}}
\and S.~Dusini\orcid{0000-0002-1128-0664}\inst{\ref{aff57}}
\and A.~Ealet\orcid{0000-0003-3070-014X}\inst{\ref{aff48}}
\and S.~Escoffier\orcid{0000-0002-2847-7498}\inst{\ref{aff58}}
\and M.~Farina\orcid{0000-0002-3089-7846}\inst{\ref{aff56}}
\and R.~Farinelli\inst{\ref{aff4}}
\and F.~Faustini\orcid{0000-0001-6274-5145}\inst{\ref{aff23},\ref{aff40}}
\and S.~Ferriol\inst{\ref{aff48}}
\and F.~Finelli\orcid{0000-0002-6694-3269}\inst{\ref{aff4},\ref{aff59}}
\and S.~Fotopoulou\orcid{0000-0002-9686-254X}\inst{\ref{aff60}}
\and M.~Frailis\orcid{0000-0002-7400-2135}\inst{\ref{aff17}}
\and E.~Franceschi\orcid{0000-0002-0585-6591}\inst{\ref{aff4}}
\and S.~Galeotta\orcid{0000-0002-3748-5115}\inst{\ref{aff17}}
\and K.~George\orcid{0000-0002-1734-8455}\inst{\ref{aff22}}
\and B.~Gillis\orcid{0000-0002-4478-1270}\inst{\ref{aff45}}
\and C.~Giocoli\orcid{0000-0002-9590-7961}\inst{\ref{aff4},\ref{aff20}}
\and P.~G\'omez-Alvarez\orcid{0000-0002-8594-5358}\inst{\ref{aff61},\ref{aff13}}
\and J.~Gracia-Carpio\inst{\ref{aff21}}
\and B.~R.~Granett\orcid{0000-0003-2694-9284}\inst{\ref{aff15}}
\and F.~Grupp\inst{\ref{aff21},\ref{aff22}}
\and S.~Gwyn\orcid{0000-0001-8221-8406}\inst{\ref{aff62}}
\and S.~V.~H.~Haugan\orcid{0000-0001-9648-7260}\inst{\ref{aff63}}
\and W.~Holmes\inst{\ref{aff64}}
\and I.~M.~Hook\orcid{0000-0002-2960-978X}\inst{\ref{aff65}}
\and F.~Hormuth\inst{\ref{aff66}}
\and A.~Hornstrup\orcid{0000-0002-3363-0936}\inst{\ref{aff67},\ref{aff68}}
\and P.~Hudelot\inst{\ref{aff69}}
\and K.~Jahnke\orcid{0000-0003-3804-2137}\inst{\ref{aff70}}
\and M.~Jhabvala\inst{\ref{aff71}}
\and E.~Keih\"anen\orcid{0000-0003-1804-7715}\inst{\ref{aff72}}
\and S.~Kermiche\orcid{0000-0002-0302-5735}\inst{\ref{aff58}}
\and A.~Kiessling\orcid{0000-0002-2590-1273}\inst{\ref{aff64}}
\and B.~Kubik\orcid{0009-0006-5823-4880}\inst{\ref{aff48}}
\and K.~Kuijken\orcid{0000-0002-3827-0175}\inst{\ref{aff35}}
\and M.~K\"ummel\orcid{0000-0003-2791-2117}\inst{\ref{aff22}}
\and M.~Kunz\orcid{0000-0002-3052-7394}\inst{\ref{aff73}}
\and H.~Kurki-Suonio\orcid{0000-0002-4618-3063}\inst{\ref{aff74},\ref{aff75}}
\and Q.~Le~Boulc'h\inst{\ref{aff76}}
\and A.~M.~C.~Le~Brun\orcid{0000-0002-0936-4594}\inst{\ref{aff77}}
\and D.~Le~Mignant\orcid{0000-0002-5339-5515}\inst{\ref{aff78}}
\and S.~Ligori\orcid{0000-0003-4172-4606}\inst{\ref{aff32}}
\and P.~B.~Lilje\orcid{0000-0003-4324-7794}\inst{\ref{aff63}}
\and V.~Lindholm\orcid{0000-0003-2317-5471}\inst{\ref{aff74},\ref{aff75}}
\and I.~Lloro\orcid{0000-0001-5966-1434}\inst{\ref{aff79}}
\and G.~Mainetti\orcid{0000-0003-2384-2377}\inst{\ref{aff76}}
\and D.~Maino\inst{\ref{aff80},\ref{aff36},\ref{aff81}}
\and E.~Maiorano\orcid{0000-0003-2593-4355}\inst{\ref{aff4}}
\and O.~Mansutti\orcid{0000-0001-5758-4658}\inst{\ref{aff17}}
\and S.~Marcin\inst{\ref{aff82}}
\and O.~Marggraf\orcid{0000-0001-7242-3852}\inst{\ref{aff83}}
\and M.~Martinelli\orcid{0000-0002-6943-7732}\inst{\ref{aff40},\ref{aff84}}
\and N.~Martinet\orcid{0000-0003-2786-7790}\inst{\ref{aff78}}
\and F.~Marulli\orcid{0000-0002-8850-0303}\inst{\ref{aff85},\ref{aff4},\ref{aff20}}
\and R.~Massey\orcid{0000-0002-6085-3780}\inst{\ref{aff86}}
\and S.~Maurogordato\inst{\ref{aff87}}
\and E.~Medinaceli\orcid{0000-0002-4040-7783}\inst{\ref{aff4}}
\and S.~Mei\orcid{0000-0002-2849-559X}\inst{\ref{aff88},\ref{aff89}}
\and M.~Melchior\inst{\ref{aff90}}
\and Y.~Mellier\inst{\ref{aff91},\ref{aff69}}
\and M.~Meneghetti\orcid{0000-0003-1225-7084}\inst{\ref{aff4},\ref{aff20}}
\and E.~Merlin\orcid{0000-0001-6870-8900}\inst{\ref{aff40}}
\and G.~Meylan\inst{\ref{aff92}}
\and A.~Mora\orcid{0000-0002-1922-8529}\inst{\ref{aff93}}
\and M.~Moresco\orcid{0000-0002-7616-7136}\inst{\ref{aff85},\ref{aff4}}
\and L.~Moscardini\orcid{0000-0002-3473-6716}\inst{\ref{aff85},\ref{aff4},\ref{aff20}}
\and R.~Nakajima\orcid{0009-0009-1213-7040}\inst{\ref{aff83}}
\and C.~Neissner\orcid{0000-0001-8524-4968}\inst{\ref{aff94},\ref{aff38}}
\and S.-M.~Niemi\inst{\ref{aff33}}
\and J.~W.~Nightingale\orcid{0000-0002-8987-7401}\inst{\ref{aff95}}
\and C.~Padilla\orcid{0000-0001-7951-0166}\inst{\ref{aff94}}
\and S.~Paltani\orcid{0000-0002-8108-9179}\inst{\ref{aff55}}
\and F.~Pasian\orcid{0000-0002-4869-3227}\inst{\ref{aff17}}
\and K.~Pedersen\inst{\ref{aff96}}
\and W.~J.~Percival\orcid{0000-0002-0644-5727}\inst{\ref{aff97},\ref{aff98},\ref{aff99}}
\and V.~Pettorino\inst{\ref{aff33}}
\and S.~Pires\orcid{0000-0002-0249-2104}\inst{\ref{aff6}}
\and G.~Polenta\orcid{0000-0003-4067-9196}\inst{\ref{aff23}}
\and M.~Poncet\inst{\ref{aff100}}
\and L.~A.~Popa\inst{\ref{aff101}}
\and L.~Pozzetti\orcid{0000-0001-7085-0412}\inst{\ref{aff4}}
\and F.~Raison\orcid{0000-0002-7819-6918}\inst{\ref{aff21}}
\and R.~Rebolo\orcid{0000-0003-3767-7085}\inst{\ref{aff44},\ref{aff102},\ref{aff103}}
\and A.~Renzi\orcid{0000-0001-9856-1970}\inst{\ref{aff1},\ref{aff57}}
\and J.~Rhodes\orcid{0000-0002-4485-8549}\inst{\ref{aff64}}
\and G.~Riccio\inst{\ref{aff27}}
\and E.~Romelli\orcid{0000-0003-3069-9222}\inst{\ref{aff17}}
\and M.~Roncarelli\orcid{0000-0001-9587-7822}\inst{\ref{aff4}}
\and E.~Rossetti\orcid{0000-0003-0238-4047}\inst{\ref{aff3}}
\and B.~Rusholme\orcid{0000-0001-7648-4142}\inst{\ref{aff104}}
\and R.~Saglia\orcid{0000-0003-0378-7032}\inst{\ref{aff22},\ref{aff21}}
\and Z.~Sakr\orcid{0000-0002-4823-3757}\inst{\ref{aff105},\ref{aff106},\ref{aff107}}
\and D.~Sapone\orcid{0000-0001-7089-4503}\inst{\ref{aff108}}
\and B.~Sartoris\orcid{0000-0003-1337-5269}\inst{\ref{aff22},\ref{aff17}}
\and J.~A.~Schewtschenko\orcid{0000-0002-4913-6393}\inst{\ref{aff45}}
\and M.~Schirmer\orcid{0000-0003-2568-9994}\inst{\ref{aff70}}
\and P.~Schneider\orcid{0000-0001-8561-2679}\inst{\ref{aff83}}
\and M.~Scodeggio\inst{\ref{aff36}}
\and A.~Secroun\orcid{0000-0003-0505-3710}\inst{\ref{aff58}}
\and G.~Seidel\orcid{0000-0003-2907-353X}\inst{\ref{aff70}}
\and S.~Serrano\orcid{0000-0002-0211-2861}\inst{\ref{aff109},\ref{aff110},\ref{aff111}}
\and P.~Simon\inst{\ref{aff83}}
\and C.~Sirignano\orcid{0000-0002-0995-7146}\inst{\ref{aff1},\ref{aff57}}
\and G.~Sirri\orcid{0000-0003-2626-2853}\inst{\ref{aff20}}
\and L.~Stanco\orcid{0000-0002-9706-5104}\inst{\ref{aff57}}
\and J.~Steinwagner\orcid{0000-0001-7443-1047}\inst{\ref{aff21}}
\and P.~Tallada-Cresp\'{i}\orcid{0000-0002-1336-8328}\inst{\ref{aff37},\ref{aff38}}
\and D.~Tavagnacco\orcid{0000-0001-7475-9894}\inst{\ref{aff17}}
\and A.~N.~Taylor\inst{\ref{aff45}}
\and H.~I.~Teplitz\orcid{0000-0002-7064-5424}\inst{\ref{aff112}}
\and I.~Tereno\inst{\ref{aff53},\ref{aff113}}
\and S.~Toft\orcid{0000-0003-3631-7176}\inst{\ref{aff11},\ref{aff114}}
\and R.~Toledo-Moreo\orcid{0000-0002-2997-4859}\inst{\ref{aff115}}
\and F.~Torradeflot\orcid{0000-0003-1160-1517}\inst{\ref{aff38},\ref{aff37}}
\and I.~Tutusaus\orcid{0000-0002-3199-0399}\inst{\ref{aff106}}
\and L.~Valenziano\orcid{0000-0002-1170-0104}\inst{\ref{aff4},\ref{aff59}}
\and J.~Valiviita\orcid{0000-0001-6225-3693}\inst{\ref{aff74},\ref{aff75}}
\and T.~Vassallo\orcid{0000-0001-6512-6358}\inst{\ref{aff22},\ref{aff17}}
\and G.~Verdoes~Kleijn\orcid{0000-0001-5803-2580}\inst{\ref{aff10}}
\and A.~Veropalumbo\orcid{0000-0003-2387-1194}\inst{\ref{aff15},\ref{aff25},\ref{aff24}}
\and Y.~Wang\orcid{0000-0002-4749-2984}\inst{\ref{aff112}}
\and J.~Weller\orcid{0000-0002-8282-2010}\inst{\ref{aff22},\ref{aff21}}
\and A.~Zacchei\orcid{0000-0003-0396-1192}\inst{\ref{aff17},\ref{aff16}}
\and G.~Zamorani\orcid{0000-0002-2318-301X}\inst{\ref{aff4}}
\and F.~M.~Zerbi\inst{\ref{aff15}}
\and I.~A.~Zinchenko\orcid{0000-0002-2944-2449}\inst{\ref{aff22}}
\and E.~Zucca\orcid{0000-0002-5845-8132}\inst{\ref{aff4}}
\and V.~Allevato\orcid{0000-0001-7232-5152}\inst{\ref{aff27}}
\and M.~Ballardini\orcid{0000-0003-4481-3559}\inst{\ref{aff116},\ref{aff117},\ref{aff4}}
\and M.~Bolzonella\orcid{0000-0003-3278-4607}\inst{\ref{aff4}}
\and E.~Bozzo\orcid{0000-0002-8201-1525}\inst{\ref{aff55}}
\and C.~Burigana\orcid{0000-0002-3005-5796}\inst{\ref{aff118},\ref{aff59}}
\and R.~Cabanac\orcid{0000-0001-6679-2600}\inst{\ref{aff106}}
\and A.~Cappi\inst{\ref{aff4},\ref{aff87}}
\and D.~Di~Ferdinando\inst{\ref{aff20}}
\and J.~A.~Escartin~Vigo\inst{\ref{aff21}}
\and L.~Gabarra\orcid{0000-0002-8486-8856}\inst{\ref{aff119}}
\and M.~Huertas-Company\orcid{0000-0002-1416-8483}\inst{\ref{aff44},\ref{aff120},\ref{aff121},\ref{aff122}}
\and J.~Mart\'{i}n-Fleitas\orcid{0000-0002-8594-569X}\inst{\ref{aff93}}
\and S.~Matthew\orcid{0000-0001-8448-1697}\inst{\ref{aff45}}
\and N.~Mauri\orcid{0000-0001-8196-1548}\inst{\ref{aff43},\ref{aff20}}
\and R.~B.~Metcalf\orcid{0000-0003-3167-2574}\inst{\ref{aff85},\ref{aff4}}
\and A.~Pezzotta\orcid{0000-0003-0726-2268}\inst{\ref{aff21}}
\and M.~P\"ontinen\orcid{0000-0001-5442-2530}\inst{\ref{aff74}}
\and C.~Porciani\orcid{0000-0002-7797-2508}\inst{\ref{aff83}}
\and I.~Risso\orcid{0000-0003-2525-7761}\inst{\ref{aff123}}
\and V.~Scottez\inst{\ref{aff91},\ref{aff124}}
\and M.~Sereno\orcid{0000-0003-0302-0325}\inst{\ref{aff4},\ref{aff20}}
\and M.~Tenti\orcid{0000-0002-4254-5901}\inst{\ref{aff20}}
\and M.~Viel\orcid{0000-0002-2642-5707}\inst{\ref{aff16},\ref{aff17},\ref{aff19},\ref{aff18},\ref{aff125}}
\and M.~Wiesmann\orcid{0009-0000-8199-5860}\inst{\ref{aff63}}
\and Y.~Akrami\orcid{0000-0002-2407-7956}\inst{\ref{aff126},\ref{aff127}}
\and I.~T.~Andika\orcid{0000-0001-6102-9526}\inst{\ref{aff128},\ref{aff129}}
\and S.~Anselmi\orcid{0000-0002-3579-9583}\inst{\ref{aff57},\ref{aff1},\ref{aff130}}
\and M.~Archidiacono\orcid{0000-0003-4952-9012}\inst{\ref{aff80},\ref{aff81}}
\and F.~Atrio-Barandela\orcid{0000-0002-2130-2513}\inst{\ref{aff131}}
\and C.~Benoist\inst{\ref{aff87}}
\and K.~Benson\inst{\ref{aff52}}
\and D.~Bertacca\orcid{0000-0002-2490-7139}\inst{\ref{aff1},\ref{aff2},\ref{aff57}}
\and M.~Bethermin\orcid{0000-0002-3915-2015}\inst{\ref{aff132}}
\and A.~Blanchard\orcid{0000-0001-8555-9003}\inst{\ref{aff106}}
\and L.~Blot\orcid{0000-0002-9622-7167}\inst{\ref{aff133},\ref{aff130}}
\and M.~L.~Brown\orcid{0000-0002-0370-8077}\inst{\ref{aff46}}
\and S.~Bruton\orcid{0000-0002-6503-5218}\inst{\ref{aff134}}
\and A.~Calabro\orcid{0000-0003-2536-1614}\inst{\ref{aff40}}
\and B.~Camacho~Quevedo\orcid{0000-0002-8789-4232}\inst{\ref{aff109},\ref{aff111}}
\and F.~Caro\inst{\ref{aff40}}
\and C.~S.~Carvalho\inst{\ref{aff113}}
\and T.~Castro\orcid{0000-0002-6292-3228}\inst{\ref{aff17},\ref{aff18},\ref{aff16},\ref{aff125}}
\and F.~Cogato\orcid{0000-0003-4632-6113}\inst{\ref{aff85},\ref{aff4}}
\and A.~R.~Cooray\orcid{0000-0002-3892-0190}\inst{\ref{aff135}}
\and O.~Cucciati\orcid{0000-0002-9336-7551}\inst{\ref{aff4}}
\and S.~Davini\orcid{0000-0003-3269-1718}\inst{\ref{aff25}}
\and F.~De~Paolis\orcid{0000-0001-6460-7563}\inst{\ref{aff136},\ref{aff137},\ref{aff138}}
\and G.~Desprez\orcid{0000-0001-8325-1742}\inst{\ref{aff10}}
\and A.~D\'iaz-S\'anchez\orcid{0000-0003-0748-4768}\inst{\ref{aff139}}
\and J.~J.~Diaz\inst{\ref{aff44}}
\and S.~Di~Domizio\orcid{0000-0003-2863-5895}\inst{\ref{aff24},\ref{aff25}}
\and J.~M.~Diego\orcid{0000-0001-9065-3926}\inst{\ref{aff140}}
\and P.-A.~Duc\orcid{0000-0003-3343-6284}\inst{\ref{aff132}}
\and Y.~Fang\inst{\ref{aff22}}
\and A.~G.~Ferrari\orcid{0009-0005-5266-4110}\inst{\ref{aff20}}
\and P.~G.~Ferreira\orcid{0000-0002-3021-2851}\inst{\ref{aff119}}
\and A.~Finoguenov\orcid{0000-0002-4606-5403}\inst{\ref{aff74}}
\and A.~Fontana\orcid{0000-0003-3820-2823}\inst{\ref{aff40}}
\and F.~Fontanot\orcid{0000-0003-4744-0188}\inst{\ref{aff17},\ref{aff16}}
\and A.~Franco\orcid{0000-0002-4761-366X}\inst{\ref{aff137},\ref{aff136},\ref{aff138}}
\and K.~Ganga\orcid{0000-0001-8159-8208}\inst{\ref{aff88}}
\and J.~Garc\'ia-Bellido\orcid{0000-0002-9370-8360}\inst{\ref{aff126}}
\and T.~Gasparetto\orcid{0000-0002-7913-4866}\inst{\ref{aff17}}
\and V.~Gautard\inst{\ref{aff141}}
\and E.~Gaztanaga\orcid{0000-0001-9632-0815}\inst{\ref{aff111},\ref{aff109},\ref{aff9}}
\and F.~Giacomini\orcid{0000-0002-3129-2814}\inst{\ref{aff20}}
\and F.~Gianotti\orcid{0000-0003-4666-119X}\inst{\ref{aff4}}
\and G.~Gozaliasl\orcid{0000-0002-0236-919X}\inst{\ref{aff142},\ref{aff74}}
\and A.~Gregorio\orcid{0000-0003-4028-8785}\inst{\ref{aff143},\ref{aff17},\ref{aff18}}
\and M.~Guidi\orcid{0000-0001-9408-1101}\inst{\ref{aff3},\ref{aff4}}
\and C.~M.~Gutierrez\orcid{0000-0001-7854-783X}\inst{\ref{aff144}}
\and A.~Hall\orcid{0000-0002-3139-8651}\inst{\ref{aff45}}
\and W.~G.~Hartley\inst{\ref{aff55}}
\and S.~Hemmati\orcid{0000-0003-2226-5395}\inst{\ref{aff104}}
\and C.~Hern\'andez-Monteagudo\orcid{0000-0001-5471-9166}\inst{\ref{aff103},\ref{aff44}}
\and H.~Hildebrandt\orcid{0000-0002-9814-3338}\inst{\ref{aff145}}
\and J.~Hjorth\orcid{0000-0002-4571-2306}\inst{\ref{aff96}}
\and J.~J.~E.~Kajava\orcid{0000-0002-3010-8333}\inst{\ref{aff146},\ref{aff147}}
\and Y.~Kang\orcid{0009-0000-8588-7250}\inst{\ref{aff55}}
\and V.~Kansal\orcid{0000-0002-4008-6078}\inst{\ref{aff148},\ref{aff149}}
\and D.~Karagiannis\orcid{0000-0002-4927-0816}\inst{\ref{aff116},\ref{aff150}}
\and K.~Kiiveri\inst{\ref{aff72}}
\and C.~C.~Kirkpatrick\inst{\ref{aff72}}
\and S.~Kruk\orcid{0000-0001-8010-8879}\inst{\ref{aff13}}
\and J.~Le~Graet\orcid{0000-0001-6523-7971}\inst{\ref{aff58}}
\and L.~Legrand\orcid{0000-0003-0610-5252}\inst{\ref{aff151},\ref{aff152}}
\and M.~Lembo\orcid{0000-0002-5271-5070}\inst{\ref{aff116},\ref{aff117}}
\and F.~Lepori\orcid{0009-0000-5061-7138}\inst{\ref{aff153}}
\and G.~Leroy\orcid{0009-0004-2523-4425}\inst{\ref{aff154},\ref{aff86}}
\and G.~F.~Lesci\orcid{0000-0002-4607-2830}\inst{\ref{aff85},\ref{aff4}}
\and J.~Lesgourgues\orcid{0000-0001-7627-353X}\inst{\ref{aff39}}
\and L.~Leuzzi\orcid{0009-0006-4479-7017}\inst{\ref{aff85},\ref{aff4}}
\and T.~I.~Liaudat\orcid{0000-0002-9104-314X}\inst{\ref{aff155}}
\and A.~Loureiro\orcid{0000-0002-4371-0876}\inst{\ref{aff156},\ref{aff157}}
\and J.~Macias-Perez\orcid{0000-0002-5385-2763}\inst{\ref{aff158}}
\and G.~Maggio\orcid{0000-0003-4020-4836}\inst{\ref{aff17}}
\and M.~Magliocchetti\orcid{0000-0001-9158-4838}\inst{\ref{aff56}}
\and E.~A.~Magnier\orcid{0000-0002-7965-2815}\inst{\ref{aff42}}
\and F.~Mannucci\orcid{0000-0002-4803-2381}\inst{\ref{aff159}}
\and R.~Maoli\orcid{0000-0002-6065-3025}\inst{\ref{aff160},\ref{aff40}}
\and C.~J.~A.~P.~Martins\orcid{0000-0002-4886-9261}\inst{\ref{aff161},\ref{aff28}}
\and L.~Maurin\orcid{0000-0002-8406-0857}\inst{\ref{aff12}}
\and M.~Miluzio\inst{\ref{aff13},\ref{aff162}}
\and P.~Monaco\orcid{0000-0003-2083-7564}\inst{\ref{aff143},\ref{aff17},\ref{aff18},\ref{aff16}}
\and C.~Moretti\orcid{0000-0003-3314-8936}\inst{\ref{aff19},\ref{aff125},\ref{aff17},\ref{aff16},\ref{aff18}}
\and G.~Morgante\inst{\ref{aff4}}
\and S.~Nadathur\orcid{0000-0001-9070-3102}\inst{\ref{aff9}}
\and K.~Naidoo\orcid{0000-0002-9182-1802}\inst{\ref{aff9}}
\and A.~Navarro-Alsina\orcid{0000-0002-3173-2592}\inst{\ref{aff83}}
\and S.~Nesseris\orcid{0000-0002-0567-0324}\inst{\ref{aff126}}
\and F.~Passalacqua\orcid{0000-0002-8606-4093}\inst{\ref{aff1},\ref{aff57}}
\and K.~Paterson\orcid{0000-0001-8340-3486}\inst{\ref{aff70}}
\and L.~Patrizii\inst{\ref{aff20}}
\and A.~Pisani\orcid{0000-0002-6146-4437}\inst{\ref{aff58},\ref{aff163}}
\and D.~Potter\orcid{0000-0002-0757-5195}\inst{\ref{aff153}}
\and S.~Quai\orcid{0000-0002-0449-8163}\inst{\ref{aff85},\ref{aff4}}
\and M.~Radovich\orcid{0000-0002-3585-866X}\inst{\ref{aff2}}
\and P.-F.~Rocci\inst{\ref{aff12}}
\and S.~Sacquegna\orcid{0000-0002-8433-6630}\inst{\ref{aff136},\ref{aff137},\ref{aff138}}
\and M.~Sahl\'en\orcid{0000-0003-0973-4804}\inst{\ref{aff164}}
\and D.~B.~Sanders\orcid{0000-0002-1233-9998}\inst{\ref{aff42}}
\and E.~Sarpa\orcid{0000-0002-1256-655X}\inst{\ref{aff19},\ref{aff125},\ref{aff18}}
\and C.~Scarlata\orcid{0000-0002-9136-8876}\inst{\ref{aff165}}
\and J.~Schaye\orcid{0000-0002-0668-5560}\inst{\ref{aff35}}
\and A.~Schneider\orcid{0000-0001-7055-8104}\inst{\ref{aff153}}
\and M.~Schultheis\inst{\ref{aff87}}
\and D.~Sciotti\orcid{0009-0008-4519-2620}\inst{\ref{aff40},\ref{aff84}}
\and E.~Sellentin\inst{\ref{aff166},\ref{aff35}}
\and F.~Shankar\orcid{0000-0001-8973-5051}\inst{\ref{aff167}}
\and L.~C.~Smith\orcid{0000-0002-3259-2771}\inst{\ref{aff168}}
\and J.~Stadel\orcid{0000-0001-7565-8622}\inst{\ref{aff153}}
\and K.~Tanidis\orcid{0000-0001-9843-5130}\inst{\ref{aff119}}
\and C.~Tao\orcid{0000-0001-7961-8177}\inst{\ref{aff58}}
\and G.~Testera\inst{\ref{aff25}}
\and R.~Teyssier\orcid{0000-0001-7689-0933}\inst{\ref{aff163}}
\and S.~Tosi\orcid{0000-0002-7275-9193}\inst{\ref{aff24},\ref{aff123}}
\and A.~Troja\orcid{0000-0003-0239-4595}\inst{\ref{aff1},\ref{aff57}}
\and M.~Tucci\inst{\ref{aff55}}
\and C.~Valieri\inst{\ref{aff20}}
\and A.~Venhola\orcid{0000-0001-6071-4564}\inst{\ref{aff169}}
\and D.~Vergani\orcid{0000-0003-0898-2216}\inst{\ref{aff4}}
\and G.~Verza\orcid{0000-0002-1886-8348}\inst{\ref{aff170}}
\and P.~Vielzeuf\orcid{0000-0003-2035-9339}\inst{\ref{aff58}}
\and N.~A.~Walton\orcid{0000-0003-3983-8778}\inst{\ref{aff168}}
\and D.~Scott\orcid{0000-0002-6878-9840}\inst{\ref{aff171}}}
										   
%%%% please do not edit the affiliation list -- contact ECEB Bureau for changes
\institute{Dipartimento di Fisica e Astronomia "G. Galilei", Universit\`a di Padova, Via Marzolo 8, 35131 Padova, Italy\label{aff1}
\and
INAF-Osservatorio Astronomico di Padova, Via dell'Osservatorio 5, 35122 Padova, Italy\label{aff2}
\and
Dipartimento di Fisica e Astronomia, Universit\`a di Bologna, Via Gobetti 93/2, 40129 Bologna, Italy\label{aff3}
\and
INAF-Osservatorio di Astrofisica e Scienza dello Spazio di Bologna, Via Piero Gobetti 93/3, 40129 Bologna, Italy\label{aff4}
\and
Jeremiah Horrocks Institute, University of Central Lancashire, Preston, PR1 2HE, UK\label{aff5}
\and
Universit\'e Paris-Saclay, Universit\'e Paris Cit\'e, CEA, CNRS, AIM, 91191, Gif-sur-Yvette, France\label{aff6}
\and
School of Physical Sciences, The Open University, Milton Keynes, MK7 6AA, UK\label{aff7}
\and
Dipartimento di Fisica e Astronomia ``G. Galilei", Universit\`a di Padova, Vicolo dell'Osservatorio 3, 35122 Padova, Italy\label{aff8}
\and
Institute of Cosmology and Gravitation, University of Portsmouth, Portsmouth PO1 3FX, UK\label{aff9}
\and
Kapteyn Astronomical Institute, University of Groningen, PO Box 800, 9700 AV Groningen, The Netherlands\label{aff10}
\and
Cosmic Dawn Center (DAWN)\label{aff11}
\and
Universit\'e Paris-Saclay, CNRS, Institut d'astrophysique spatiale, 91405, Orsay, France\label{aff12}
\and
ESAC/ESA, Camino Bajo del Castillo, s/n., Urb. Villafranca del Castillo, 28692 Villanueva de la Ca\~nada, Madrid, Spain\label{aff13}
\and
School of Mathematics and Physics, University of Surrey, Guildford, Surrey, GU2 7XH, UK\label{aff14}
\and
INAF-Osservatorio Astronomico di Brera, Via Brera 28, 20122 Milano, Italy\label{aff15}
\and
IFPU, Institute for Fundamental Physics of the Universe, via Beirut 2, 34151 Trieste, Italy\label{aff16}
\and
INAF-Osservatorio Astronomico di Trieste, Via G. B. Tiepolo 11, 34143 Trieste, Italy\label{aff17}
\and
INFN, Sezione di Trieste, Via Valerio 2, 34127 Trieste TS, Italy\label{aff18}
\and
SISSA, International School for Advanced Studies, Via Bonomea 265, 34136 Trieste TS, Italy\label{aff19}
\and
INFN-Sezione di Bologna, Viale Berti Pichat 6/2, 40127 Bologna, Italy\label{aff20}
\and
Max Planck Institute for Extraterrestrial Physics, Giessenbachstr. 1, 85748 Garching, Germany\label{aff21}
\and
Universit\"ats-Sternwarte M\"unchen, Fakult\"at f\"ur Physik, Ludwig-Maximilians-Universit\"at M\"unchen, Scheinerstrasse 1, 81679 M\"unchen, Germany\label{aff22}
\and
Space Science Data Center, Italian Space Agency, via del Politecnico snc, 00133 Roma, Italy\label{aff23}
\and
Dipartimento di Fisica, Universit\`a di Genova, Via Dodecaneso 33, 16146, Genova, Italy\label{aff24}
\and
INFN-Sezione di Genova, Via Dodecaneso 33, 16146, Genova, Italy\label{aff25}
\and
Department of Physics "E. Pancini", University Federico II, Via Cinthia 6, 80126, Napoli, Italy\label{aff26}
\and
INAF-Osservatorio Astronomico di Capodimonte, Via Moiariello 16, 80131 Napoli, Italy\label{aff27}
\and
Instituto de Astrof\'isica e Ci\^encias do Espa\c{c}o, Universidade do Porto, CAUP, Rua das Estrelas, PT4150-762 Porto, Portugal\label{aff28}
\and
Faculdade de Ci\^encias da Universidade do Porto, Rua do Campo de Alegre, 4150-007 Porto, Portugal\label{aff29}
\and
Dipartimento di Fisica, Universit\`a degli Studi di Torino, Via P. Giuria 1, 10125 Torino, Italy\label{aff30}
\and
INFN-Sezione di Torino, Via P. Giuria 1, 10125 Torino, Italy\label{aff31}
\and
INAF-Osservatorio Astrofisico di Torino, Via Osservatorio 20, 10025 Pino Torinese (TO), Italy\label{aff32}
\and
European Space Agency/ESTEC, Keplerlaan 1, 2201 AZ Noordwijk, The Netherlands\label{aff33}
\and
Institute Lorentz, Leiden University, Niels Bohrweg 2, 2333 CA Leiden, The Netherlands\label{aff34}
\and
Leiden Observatory, Leiden University, Einsteinweg 55, 2333 CC Leiden, The Netherlands\label{aff35}
\and
INAF-IASF Milano, Via Alfonso Corti 12, 20133 Milano, Italy\label{aff36}
\and
Centro de Investigaciones Energ\'eticas, Medioambientales y Tecnol\'ogicas (CIEMAT), Avenida Complutense 40, 28040 Madrid, Spain\label{aff37}
\and
Port d'Informaci\'{o} Cient\'{i}fica, Campus UAB, C. Albareda s/n, 08193 Bellaterra (Barcelona), Spain\label{aff38}
\and
Institute for Theoretical Particle Physics and Cosmology (TTK), RWTH Aachen University, 52056 Aachen, Germany\label{aff39}
\and
INAF-Osservatorio Astronomico di Roma, Via Frascati 33, 00078 Monteporzio Catone, Italy\label{aff40}
\and
INFN section of Naples, Via Cinthia 6, 80126, Napoli, Italy\label{aff41}
\and
Institute for Astronomy, University of Hawaii, 2680 Woodlawn Drive, Honolulu, HI 96822, USA\label{aff42}
\and
Dipartimento di Fisica e Astronomia "Augusto Righi" - Alma Mater Studiorum Universit\`a di Bologna, Viale Berti Pichat 6/2, 40127 Bologna, Italy\label{aff43}
\and
Instituto de Astrof\'{\i}sica de Canarias, V\'{\i}a L\'actea, 38205 La Laguna, Tenerife, Spain\label{aff44}
\and
Institute for Astronomy, University of Edinburgh, Royal Observatory, Blackford Hill, Edinburgh EH9 3HJ, UK\label{aff45}
\and
Jodrell Bank Centre for Astrophysics, Department of Physics and Astronomy, University of Manchester, Oxford Road, Manchester M13 9PL, UK\label{aff46}
\and
European Space Agency/ESRIN, Largo Galileo Galilei 1, 00044 Frascati, Roma, Italy\label{aff47}
\and
Universit\'e Claude Bernard Lyon 1, CNRS/IN2P3, IP2I Lyon, UMR 5822, Villeurbanne, F-69100, France\label{aff48}
\and
Institut de Ci\`{e}ncies del Cosmos (ICCUB), Universitat de Barcelona (IEEC-UB), Mart\'{i} i Franqu\`{e}s 1, 08028 Barcelona, Spain\label{aff49}
\and
Instituci\'o Catalana de Recerca i Estudis Avan\c{c}ats (ICREA), Passeig de Llu\'{\i}s Companys 23, 08010 Barcelona, Spain\label{aff50}
\and
UCB Lyon 1, CNRS/IN2P3, IUF, IP2I Lyon, 4 rue Enrico Fermi, 69622 Villeurbanne, France\label{aff51}
\and
Mullard Space Science Laboratory, University College London, Holmbury St Mary, Dorking, Surrey RH5 6NT, UK\label{aff52}
\and
Departamento de F\'isica, Faculdade de Ci\^encias, Universidade de Lisboa, Edif\'icio C8, Campo Grande, PT1749-016 Lisboa, Portugal\label{aff53}
\and
Instituto de Astrof\'isica e Ci\^encias do Espa\c{c}o, Faculdade de Ci\^encias, Universidade de Lisboa, Campo Grande, 1749-016 Lisboa, Portugal\label{aff54}
\and
Department of Astronomy, University of Geneva, ch. d'Ecogia 16, 1290 Versoix, Switzerland\label{aff55}
\and
INAF-Istituto di Astrofisica e Planetologia Spaziali, via del Fosso del Cavaliere, 100, 00100 Roma, Italy\label{aff56}
\and
INFN-Padova, Via Marzolo 8, 35131 Padova, Italy\label{aff57}
\and
Aix-Marseille Universit\'e, CNRS/IN2P3, CPPM, Marseille, France\label{aff58}
\and
INFN-Bologna, Via Irnerio 46, 40126 Bologna, Italy\label{aff59}
\and
School of Physics, HH Wills Physics Laboratory, University of Bristol, Tyndall Avenue, Bristol, BS8 1TL, UK\label{aff60}
\and
FRACTAL S.L.N.E., calle Tulip\'an 2, Portal 13 1A, 28231, Las Rozas de Madrid, Spain\label{aff61}
\and
NRC Herzberg, 5071 West Saanich Rd, Victoria, BC V9E 2E7, Canada\label{aff62}
\and
Institute of Theoretical Astrophysics, University of Oslo, P.O. Box 1029 Blindern, 0315 Oslo, Norway\label{aff63}
\and
Jet Propulsion Laboratory, California Institute of Technology, 4800 Oak Grove Drive, Pasadena, CA, 91109, USA\label{aff64}
\and
Department of Physics, Lancaster University, Lancaster, LA1 4YB, UK\label{aff65}
\and
Felix Hormuth Engineering, Goethestr. 17, 69181 Leimen, Germany\label{aff66}
\and
Technical University of Denmark, Elektrovej 327, 2800 Kgs. Lyngby, Denmark\label{aff67}
\and
Cosmic Dawn Center (DAWN), Denmark\label{aff68}
\and
Institut d'Astrophysique de Paris, UMR 7095, CNRS, and Sorbonne Universit\'e, 98 bis boulevard Arago, 75014 Paris, France\label{aff69}
\and
Max-Planck-Institut f\"ur Astronomie, K\"onigstuhl 17, 69117 Heidelberg, Germany\label{aff70}
\and
NASA Goddard Space Flight Center, Greenbelt, MD 20771, USA\label{aff71}
\and
Department of Physics and Helsinki Institute of Physics, Gustaf H\"allstr\"omin katu 2, 00014 University of Helsinki, Finland\label{aff72}
\and
Universit\'e de Gen\`eve, D\'epartement de Physique Th\'eorique and Centre for Astroparticle Physics, 24 quai Ernest-Ansermet, CH-1211 Gen\`eve 4, Switzerland\label{aff73}
\and
Department of Physics, P.O. Box 64, 00014 University of Helsinki, Finland\label{aff74}
\and
Helsinki Institute of Physics, Gustaf H{\"a}llstr{\"o}min katu 2, University of Helsinki, Helsinki, Finland\label{aff75}
\and
Centre de Calcul de l'IN2P3/CNRS, 21 avenue Pierre de Coubertin 69627 Villeurbanne Cedex, France\label{aff76}
\and
Laboratoire d'etude de l'Univers et des phenomenes eXtremes, Observatoire de Paris, Universit\'e PSL, Sorbonne Universit\'e, CNRS, 92190 Meudon, France\label{aff77}
\and
Aix-Marseille Universit\'e, CNRS, CNES, LAM, Marseille, France\label{aff78}
\and
SKA Observatory, Jodrell Bank, Lower Withington, Macclesfield, Cheshire SK11 9FT, UK\label{aff79}
\and
Dipartimento di Fisica "Aldo Pontremoli", Universit\`a degli Studi di Milano, Via Celoria 16, 20133 Milano, Italy\label{aff80}
\and
INFN-Sezione di Milano, Via Celoria 16, 20133 Milano, Italy\label{aff81}
\and
University of Applied Sciences and Arts of Northwestern Switzerland, School of Computer Science, 5210 Windisch, Switzerland\label{aff82}
\and
Universit\"at Bonn, Argelander-Institut f\"ur Astronomie, Auf dem H\"ugel 71, 53121 Bonn, Germany\label{aff83}
\and
INFN-Sezione di Roma, Piazzale Aldo Moro, 2 - c/o Dipartimento di Fisica, Edificio G. Marconi, 00185 Roma, Italy\label{aff84}
\and
Dipartimento di Fisica e Astronomia "Augusto Righi" - Alma Mater Studiorum Universit\`a di Bologna, via Piero Gobetti 93/2, 40129 Bologna, Italy\label{aff85}
\and
Department of Physics, Institute for Computational Cosmology, Durham University, South Road, Durham, DH1 3LE, UK\label{aff86}
\and
Universit\'e C\^{o}te d'Azur, Observatoire de la C\^{o}te d'Azur, CNRS, Laboratoire Lagrange, Bd de l'Observatoire, CS 34229, 06304 Nice cedex 4, France\label{aff87}
\and
Universit\'e Paris Cit\'e, CNRS, Astroparticule et Cosmologie, 75013 Paris, France\label{aff88}
\and
CNRS-UCB International Research Laboratory, Centre Pierre Binetruy, IRL2007, CPB-IN2P3, Berkeley, USA\label{aff89}
\and
University of Applied Sciences and Arts of Northwestern Switzerland, School of Engineering, 5210 Windisch, Switzerland\label{aff90}
\and
Institut d'Astrophysique de Paris, 98bis Boulevard Arago, 75014, Paris, France\label{aff91}
\and
Institute of Physics, Laboratory of Astrophysics, Ecole Polytechnique F\'ed\'erale de Lausanne (EPFL), Observatoire de Sauverny, 1290 Versoix, Switzerland\label{aff92}
\and
Aurora Technology for European Space Agency (ESA), Camino bajo del Castillo, s/n, Urbanizacion Villafranca del Castillo, Villanueva de la Ca\~nada, 28692 Madrid, Spain\label{aff93}
\and
Institut de F\'{i}sica d'Altes Energies (IFAE), The Barcelona Institute of Science and Technology, Campus UAB, 08193 Bellaterra (Barcelona), Spain\label{aff94}
\and
School of Mathematics, Statistics and Physics, Newcastle University, Herschel Building, Newcastle-upon-Tyne, NE1 7RU, UK\label{aff95}
\and
DARK, Niels Bohr Institute, University of Copenhagen, Jagtvej 155, 2200 Copenhagen, Denmark\label{aff96}
\and
Waterloo Centre for Astrophysics, University of Waterloo, Waterloo, Ontario N2L 3G1, Canada\label{aff97}
\and
Department of Physics and Astronomy, University of Waterloo, Waterloo, Ontario N2L 3G1, Canada\label{aff98}
\and
Perimeter Institute for Theoretical Physics, Waterloo, Ontario N2L 2Y5, Canada\label{aff99}
\and
Centre National d'Etudes Spatiales -- Centre spatial de Toulouse, 18 avenue Edouard Belin, 31401 Toulouse Cedex 9, France\label{aff100}
\and
Institute of Space Science, Str. Atomistilor, nr. 409 M\u{a}gurele, Ilfov, 077125, Romania\label{aff101}
\and
Consejo Superior de Investigaciones Cientificas, Calle Serrano 117, 28006 Madrid, Spain\label{aff102}
\and
Universidad de La Laguna, Departamento de Astrof\'{\i}sica, 38206 La Laguna, Tenerife, Spain\label{aff103}
\and
Caltech/IPAC, 1200 E. California Blvd., Pasadena, CA 91125, USA\label{aff104}
\and
Institut f\"ur Theoretische Physik, University of Heidelberg, Philosophenweg 16, 69120 Heidelberg, Germany\label{aff105}
\and
Institut de Recherche en Astrophysique et Plan\'etologie (IRAP), Universit\'e de Toulouse, CNRS, UPS, CNES, 14 Av. Edouard Belin, 31400 Toulouse, France\label{aff106}
\and
Universit\'e St Joseph; Faculty of Sciences, Beirut, Lebanon\label{aff107}
\and
Departamento de F\'isica, FCFM, Universidad de Chile, Blanco Encalada 2008, Santiago, Chile\label{aff108}
\and
Institut d'Estudis Espacials de Catalunya (IEEC),  Edifici RDIT, Campus UPC, 08860 Castelldefels, Barcelona, Spain\label{aff109}
\and
Satlantis, University Science Park, Sede Bld 48940, Leioa-Bilbao, Spain\label{aff110}
\and
Institute of Space Sciences (ICE, CSIC), Campus UAB, Carrer de Can Magrans, s/n, 08193 Barcelona, Spain\label{aff111}
\and
Infrared Processing and Analysis Center, California Institute of Technology, Pasadena, CA 91125, USA\label{aff112}
\and
Instituto de Astrof\'isica e Ci\^encias do Espa\c{c}o, Faculdade de Ci\^encias, Universidade de Lisboa, Tapada da Ajuda, 1349-018 Lisboa, Portugal\label{aff113}
\and
Niels Bohr Institute, University of Copenhagen, Jagtvej 128, 2200 Copenhagen, Denmark\label{aff114}
\and
Universidad Polit\'ecnica de Cartagena, Departamento de Electr\'onica y Tecnolog\'ia de Computadoras,  Plaza del Hospital 1, 30202 Cartagena, Spain\label{aff115}
\and
Dipartimento di Fisica e Scienze della Terra, Universit\`a degli Studi di Ferrara, Via Giuseppe Saragat 1, 44122 Ferrara, Italy\label{aff116}
\and
Istituto Nazionale di Fisica Nucleare, Sezione di Ferrara, Via Giuseppe Saragat 1, 44122 Ferrara, Italy\label{aff117}
\and
INAF, Istituto di Radioastronomia, Via Piero Gobetti 101, 40129 Bologna, Italy\label{aff118}
\and
Department of Physics, Oxford University, Keble Road, Oxford OX1 3RH, UK\label{aff119}
\and
Instituto de Astrof\'isica de Canarias (IAC); Departamento de Astrof\'isica, Universidad de La Laguna (ULL), 38200, La Laguna, Tenerife, Spain\label{aff120}
\and
Universit\'e PSL, Observatoire de Paris, Sorbonne Universit\'e, CNRS, LERMA, 75014, Paris, France\label{aff121}
\and
Universit\'e Paris-Cit\'e, 5 Rue Thomas Mann, 75013, Paris, France\label{aff122}
\and
INAF-Osservatorio Astronomico di Brera, Via Brera 28, 20122 Milano, Italy, and INFN-Sezione di Genova, Via Dodecaneso 33, 16146, Genova, Italy\label{aff123}
\and
ICL, Junia, Universit\'e Catholique de Lille, LITL, 59000 Lille, France\label{aff124}
\and
ICSC - Centro Nazionale di Ricerca in High Performance Computing, Big Data e Quantum Computing, Via Magnanelli 2, Bologna, Italy\label{aff125}
\and
Instituto de F\'isica Te\'orica UAM-CSIC, Campus de Cantoblanco, 28049 Madrid, Spain\label{aff126}
\and
CERCA/ISO, Department of Physics, Case Western Reserve University, 10900 Euclid Avenue, Cleveland, OH 44106, USA\label{aff127}
\and
Technical University of Munich, TUM School of Natural Sciences, Physics Department, James-Franck-Str.~1, 85748 Garching, Germany\label{aff128}
\and
Max-Planck-Institut f\"ur Astrophysik, Karl-Schwarzschild-Str.~1, 85748 Garching, Germany\label{aff129}
\and
Laboratoire Univers et Th\'eorie, Observatoire de Paris, Universit\'e PSL, Universit\'e Paris Cit\'e, CNRS, 92190 Meudon, France\label{aff130}
\and
Departamento de F{\'\i}sica Fundamental. Universidad de Salamanca. Plaza de la Merced s/n. 37008 Salamanca, Spain\label{aff131}
\and
Universit\'e de Strasbourg, CNRS, Observatoire astronomique de Strasbourg, UMR 7550, 67000 Strasbourg, France\label{aff132}
\and
Center for Data-Driven Discovery, Kavli IPMU (WPI), UTIAS, The University of Tokyo, Kashiwa, Chiba 277-8583, Japan\label{aff133}
\and
California Institute of Technology, 1200 E California Blvd, Pasadena, CA 91125, USA\label{aff134}
\and
Department of Physics \& Astronomy, University of California Irvine, Irvine CA 92697, USA\label{aff135}
\and
Department of Mathematics and Physics E. De Giorgi, University of Salento, Via per Arnesano, CP-I93, 73100, Lecce, Italy\label{aff136}
\and
INFN, Sezione di Lecce, Via per Arnesano, CP-193, 73100, Lecce, Italy\label{aff137}
\and
INAF-Sezione di Lecce, c/o Dipartimento Matematica e Fisica, Via per Arnesano, 73100, Lecce, Italy\label{aff138}
\and
Departamento F\'isica Aplicada, Universidad Polit\'ecnica de Cartagena, Campus Muralla del Mar, 30202 Cartagena, Murcia, Spain\label{aff139}
\and
Instituto de F\'isica de Cantabria, Edificio Juan Jord\'a, Avenida de los Castros, 39005 Santander, Spain\label{aff140}
\and
CEA Saclay, DFR/IRFU, Service d'Astrophysique, Bat. 709, 91191 Gif-sur-Yvette, France\label{aff141}
\and
Department of Computer Science, Aalto University, PO Box 15400, Espoo, FI-00 076, Finland\label{aff142}
\and
Dipartimento di Fisica - Sezione di Astronomia, Universit\`a di Trieste, Via Tiepolo 11, 34131 Trieste, Italy\label{aff143}
\and
Instituto de Astrof\'\i sica de Canarias, c/ Via Lactea s/n, La Laguna 38200, Spain. Departamento de Astrof\'\i sica de la Universidad de La Laguna, Avda. Francisco Sanchez, La Laguna, 38200, Spain\label{aff144}
\and
Ruhr University Bochum, Faculty of Physics and Astronomy, Astronomical Institute (AIRUB), German Centre for Cosmological Lensing (GCCL), 44780 Bochum, Germany\label{aff145}
\and
Department of Physics and Astronomy, Vesilinnantie 5, 20014 University of Turku, Finland\label{aff146}
\and
Serco for European Space Agency (ESA), Camino bajo del Castillo, s/n, Urbanizacion Villafranca del Castillo, Villanueva de la Ca\~nada, 28692 Madrid, Spain\label{aff147}
\and
ARC Centre of Excellence for Dark Matter Particle Physics, Melbourne, Australia\label{aff148}
\and
Centre for Astrophysics \& Supercomputing, Swinburne University of Technology,  Hawthorn, Victoria 3122, Australia\label{aff149}
\and
Department of Physics and Astronomy, University of the Western Cape, Bellville, Cape Town, 7535, South Africa\label{aff150}
\and
DAMTP, Centre for Mathematical Sciences, Wilberforce Road, Cambridge CB3 0WA, UK\label{aff151}
\and
Kavli Institute for Cosmology Cambridge, Madingley Road, Cambridge, CB3 0HA, UK\label{aff152}
\and
Department of Astrophysics, University of Zurich, Winterthurerstrasse 190, 8057 Zurich, Switzerland\label{aff153}
\and
Department of Physics, Centre for Extragalactic Astronomy, Durham University, South Road, Durham, DH1 3LE, UK\label{aff154}
\and
IRFU, CEA, Universit\'e Paris-Saclay 91191 Gif-sur-Yvette Cedex, France\label{aff155}
\and
Oskar Klein Centre for Cosmoparticle Physics, Department of Physics, Stockholm University, Stockholm, SE-106 91, Sweden\label{aff156}
\and
Astrophysics Group, Blackett Laboratory, Imperial College London, London SW7 2AZ, UK\label{aff157}
\and
Univ. Grenoble Alpes, CNRS, Grenoble INP, LPSC-IN2P3, 53, Avenue des Martyrs, 38000, Grenoble, France\label{aff158}
\and
INAF-Osservatorio Astrofisico di Arcetri, Largo E. Fermi 5, 50125, Firenze, Italy\label{aff159}
\and
Dipartimento di Fisica, Sapienza Universit\`a di Roma, Piazzale Aldo Moro 2, 00185 Roma, Italy\label{aff160}
\and
Centro de Astrof\'{\i}sica da Universidade do Porto, Rua das Estrelas, 4150-762 Porto, Portugal\label{aff161}
\and
HE Space for European Space Agency (ESA), Camino bajo del Castillo, s/n, Urbanizacion Villafranca del Castillo, Villanueva de la Ca\~nada, 28692 Madrid, Spain\label{aff162}
\and
Department of Astrophysical Sciences, Peyton Hall, Princeton University, Princeton, NJ 08544, USA\label{aff163}
\and
Theoretical astrophysics, Department of Physics and Astronomy, Uppsala University, Box 515, 751 20 Uppsala, Sweden\label{aff164}
\and
Minnesota Institute for Astrophysics, University of Minnesota, 116 Church St SE, Minneapolis, MN 55455, USA\label{aff165}
\and
Mathematical Institute, University of Leiden, Einsteinweg 55, 2333 CA Leiden, The Netherlands\label{aff166}
\and
School of Physics \& Astronomy, University of Southampton, Highfield Campus, Southampton SO17 1BJ, UK\label{aff167}
\and
Institute of Astronomy, University of Cambridge, Madingley Road, Cambridge CB3 0HA, UK\label{aff168}
\and
Space physics and astronomy research unit, University of Oulu, Pentti Kaiteran katu 1, FI-90014 Oulu, Finland\label{aff169}
\and
Center for Computational Astrophysics, Flatiron Institute, 162 5th Avenue, 10010, New York, NY, USA\label{aff170}
\and
Department of Physics and Astronomy, University of British Columbia, Vancouver, BC V6T 1Z1, Canada\label{aff171}}    
% 
% Put your abstract here:
%
\abstract{Our understanding of cosmic star formation at $z > 3$ used to largely rely on rest-frame UV observations. However, these observations overlook dusty and massive sources, resulting in an incomplete census of early star-forming galaxies. Recently, infrared data from \textit{Spitzer} and the \textit{James Webb} Space Telescope (JWST) have revealed a hidden population at $z\sim3$--$6$ with extreme red colours. 
Taking advantage of the overlap between imaging in the Euclid Deep Fields (EDFs), covering about $60$ deg$^2$, and ancillary \textit{Spitzer} observations, we identified $27\,000$ extremely red objects with $\HE - \mathrm{IRAC2} > 2.25$ (dubbed HIEROs) down to a $10\,\sigma$ completeness magnitude limit of $\mathrm{IRAC2} = 22.5$ AB. After a visual investigation to discard artefacts and any objects with troubling photometry, we ended up with a final sample of $3\,900$ candidates. We retrieved the physical parameter estimates for these objects from the SED-fitting tool \texttt{CIGALE}. 
Our results confirm that HIERO galaxies may populate the high-mass end of the stellar mass function at $z>3$, with some sources reaching extreme stellar masses ($M_*>10^{11}M_\odot$) and exhibiting high dust attenuation values ($A_V>3$). However, we consider the stellar mass estimates unreliable for sources at $z>3.5$. For this reason, we favour a more conservative lower-$z$ solution. The challenges faced by SED-fitting tools in accurately characterising these objects underscore the need for further studies, incorporating both observations at shorter wavelengths and spectroscopic data. \Euclid spectra will help resolve degeneracies and better constrain the physical properties of the brightest galaxies. Given the extreme nature of this population, characterising these sources is crucial for building a comprehensive picture of galaxy evolution and stellar mass assembly across most of the history of the Universe. This work demonstrates \Euclid's potential to provide statistical samples of rare objects, such as massive, dust-obscured galaxies at $z>3$, which will be prime targets for JWST, the Atacama Large Millimeter/Submillimeter Array (ALMA) and the Extremely Large Telescope (ELT).
}
%
% Provide up to five key words:
%
    \keywords{methods: observational -- techniques: photometric -- galaxies: evolution -- galaxies: high-redshift -- infrared: galaxies}
%    from the list in
%     https://www.aanda.org/for-authors/latex-issues/information-files#pop}
%
% Add short versions of title and author list for page headings
%
   \titlerunning{A Q$1$ investigation of red objects in EDFs}
   %\authorrunning{Euclid Consortium: Girardi et al.}
   
   \maketitle
%
%-------------------------------------------------------------------
%
%
%   Start the main text of your paper here
%
   
\section{\label{sc:Intro}Introduction}
%
%Understanding the evolution of galaxies across cosmic time has always been an important goal for observational extragalactic astronomy. Observational results are indeed essential to check our theoretical predictions, which must reproduce our data.
Understanding the evolution of galaxies throughout cosmic history has always been a fundamental objective of extragalactic astronomy. Observational constraints are crucial for validating theoretical predictions, which must be able to accurately replicate empirical phenomena. A key challenge in this endeavour lies in achieving a complete census of the galaxy population, particularly during the early epochs of the Universe.

The \HST (HST) has been instrumental in the study of high-redshift galaxies, primarily through observations of their rest-frame ultraviolet (UV) emission. These kinds of galaxies have been extensively characterised across the redshift range $3 \lesssim z \lesssim 11$ \citep{steidel1993deep, steidel1995lyman, madau1996high, steidel1999lyman, bouwens2015uv, Oesch2016remarkably}. However, this UV-based selection systematically misses massive, dusty galaxies, underestimating the true stellar mass function (MF) at different epochs \citep{rodighiero2007unveiling,wang2019dominant}. Such dust-obscured sources are faint or undetected even in the deepest HST observations, earning them the label of `HST-dark galaxies' or `optically-dark galaxies' (ODGs).

Longer-wavelength facilities such as \textit{Spitzer} and the Atacama Large Millimeter/Submillimiter Array (ALMA) have played a crucial role in unveiling this population \citep{caputi2015spitzer, franco2018goods, dudzevivciute2020alma}. Recently, the \textit{James Webb} Space Telescope (JWST) has revolutionised the field by providing robust photometric redshifts and stellar mass estimates for these galaxies, thanks to its sensitivity and spatial resolution in the near-infrared \citep{gardner2023james, barrufet2023unveiling, rodighiero2023jwst}. Nonetheless, JWST's relatively narrow field of view limits its capacity for statistical investigations, which are essential for understanding the broader implications of this population on galaxy evolution.

The \Euclid mission \citep{laureijs2011euclid, EuclidSkyOverview}, launched in July 2023 by the European Space Agency (ESA), provides an unparalleled opportunity to address these limitations. Equipped with the Visible Camera (VIS) for optical imaging \citep{EuclidSkyVIS} and the Near-Infrared Spectrometer and Photometer (NISP) for near-infrared observations \citep{Schirmer-EP29, EuclidSkyNISP}, \Euclid is optimised for wide-field surveys, enabling statistical analyses of rare galaxy populations. With its Early Release Observations \citep[EROs;][]{EROcite}, \Euclid has already demonstrated its potential for identifying massive, dusty galaxies, especially when combined with ancillary data from \textit{Spitzer}/Infrared Array Camera (IRAC; Girardi et al. in prep.). 

With the first quick data release \citep[Q$1$;][]{Q1cite}, we now have access to high-quality observations covering an area of approximately $60\,\mathrm{deg}^2$ area. This study focuses on a specific subset of ODGs known as HST-to-IRAC Extremely Red Objects (HIEROs), identified using the colour criterion $\HE-\mathrm{IRAC2} > 2.25$ \citep{wang2016infrared, Caputi12}. By leveraging the overlap between EDFs and existing \textit{Spitzer} imaging, we aim to refine the photometric redshifts and stellar mass estimates of this population. We are particularly interested in this population due to their contribution to the high-mass end of the stellar mass function at $z \gtrsim 4$ \citep{barrufet2023unveiling, rodighiero2023jwst, wang2024true, traina2024a3cosmos, rodighiero2007unveiling, gottumukkala2024unveiling}. 

The paper is structured as follows: in Sect. \ref{sc:data} we describe the \Euclid data products and the creation of the \textit{Spitzer}/IRAC photometric catalogue; in Sect. \ref{sc:methods} we explain the selection of our sample and the methodology used to derive their physical properties; and in Sect. \ref{sc:Results}, we present our statistical analysis and discuss the results.

Throughout this work, we adopt a $\Lambda$CDM cosmology with parameters from \cite{Planck2016} and a \cite{chabrier2003galactic} Initial Mass Function (IMF). All magnitudes are reported in the AB system.

\section{Data description} \label{sc:data}
\subsection{\Euclid catalogues}
We exploited the official catalogues released inside the Euclid Consortium for Q$1$. The detailed description of the data can be found in \cite{Q1-TP001}, \cite{Q1-TP002}, \cite{Q1-TP003}, and \cite{Q1-TP004}. 

In summary, Q$1$ extragalactic observations cover $63\,\mathrm{deg}^2$ across three fields: the Euclid Deep Field Fornax (EDF-F), covering $12\,\mathrm{deg}^2$; the Euclid Deep Field North (EDF-N), with $22\,\mathrm{deg}^2$; and the Euclid Deep Field South (EDF-S), with $28\,\mathrm{deg}^2$. All the fields have been observed in the four \Euclid bands, covering from the visible \citep[$\IE$,][]{EuclidSkyVIS} to the near-infrared \citep[NISP, $\YE$, $\JE$, and $\HE$ band, see][]{EuclidSkyNISP}. These space-based observations are further supplemented by ground-based data collected with different instruments, covering wavelengths from $0.3\,\micron$ to $0.9\,\micron$. The latter are included in the officially released dataset as part of the Ultraviolet Near-Infrared Optical Northern Survey (UNIONS, Gwyn et al. in prep.) or the Dark Energy Survey \citep{DES}. The available bands are reported in Table \ref{tab:filters}. 

\subsection{\textit{Spitzer}/IRAC photometry}
\label{IRAC}
We analysed IRAC images\footnote{Retrieved at this page: \url{https://exchg.calet.org/Spitzer/linear/}} described in \cite{Moneti-EP17}. These observations partially overlap with the EDFs and are part of the Cosmic Dawn survey \citep{EP-McPartland}. It is worth noting that the images do not have uniform coverage, both in area and depth, since they were collected by different programmes.

For consistency, we adopt the same IRAC photometry included in the pipeline used to derive the physical parameters for the \Euclid sources \citep{Q1-SP031}. We exploited the two available IRAC bands at $3.6\,\micron$ (IRAC1) and $4.5\,\micron$  (IRAC2). In this work we do not consider the photometry at $5.8\,\micron$ (IRAC3) and $8.0\,\micron$ (IRAC4), since that is much shallower and inhomogeneous with respect to the other bands (see Table~\ref{tab:filters}).

While more details on the IRAC photometry measurements can be found in \citet{Q1-SP011}, in the following we briefly summarise the methodology.
Using the \texttt{photutils} Python package \citep{larry_bradley_2024_10967176}, we subtracted the background from each image by calculating the median with a $3\times3$ pixel filter. Again with \texttt{photutils}, aperture photometry was then performed on the IRAC images, forcing the detection to the \Euclid source positions and using a $1\arcsecond$ radius aperture. 

This choice excludes any source that would be detectable by IRAC but not by the \Euclid bands.
%However, given the different PSF of the two instruments, which leads to a sever problem of blending sources in the IRAC images, and in line with the idea of achieve not a complete, but a reliable sample, we decided to not consider this kind of objects for this paper.
However, due to the different PSFs of the two instruments, which cause significant source blending in the IRAC images, and in line with our goal of obtaining a reliable rather than a complete sample, we chose not to include in this study objects presenting blending or contamination effects.
When available, future works will exploit the official \Euclid catalogue where the debleding has been performed on the \textit{Spitzer} images.

For EDF-N, we conducted a separate extraction using co-added IRAC1 and IRAC2 \textit{Spitzer} images, weighted by uncertainty maps, and measured fluxes within Kron apertures \citep{kron}. We applied a Kron scaling factor of $1.8$ and a minimum unscaled radius of $2.5$ pixels. Aperture and Kron fluxes from the separate extractions were compared to derive aperture-to-total corrections, which were uniformly applied to all filters. The final fluxes are consistent with the catalogues described in \cite{EP-Zalesky}, which cover two of the three EDFs. 
%The more detailed methodology can be found in \citet{Q1-SP011}.

\section{Methods \label{sc:methods}  }
\subsection{HIERO sample selection} \label{HIERO selection}
We produced a merged \Euclid $+$ \textit{Spitzer} catalogue by matching the IDs of the sources, given the application of the photometry performed on the \textit{Spitzer} images at the \Euclid positions. The available bands and the respective observed depths are reported in Table~\ref{tab:filters}.

\begin{table}[]
    \centering
    \caption{All the available bands in the Q$1$ data release, differentiating the three fields. The reported magnitudes are the $10\,\sigma$ observed depths. Optical and \Euclid magnitudes refer to an extended source in an aperture diameter of twice the FWHM. For IRAC IRAC1 and IRAC2 values see \cite{Moneti-EP17} and \cite{EP-McPartland}. For IRAC IRAC3 and IRAC4 we report the depth derived from the IRAC catalogue, after correcting from aperture magnitude to total magnitude.}
    \label{tab:filters}
    \begin{tabular}{lccccc}
    \hline
    \hline
    \noalign{\vskip 2pt}
        Band &  $\lambda_{\rm eff}$ [\micron] & EDF-F & EDF-N & EDF-S \\
        \noalign{\vskip 2pt}
        \hline
        \noalign{\vskip 2pt}
        CFHT/MegaCam $u$      & $0.372$ & \dots  & $23.5$ & \dots \\
        HSC $g$               & $0.480$ & \dots & $25.3$ & \dots \\
        CFHT/MegaCam $r$      & $0.640$ & \dots  & $24.1$ & \dots \\
        PAN-STARRS $i$        & $0.755$ & \dots  & $23.3$ & \dots \\
        HSC $z$               & $0.891$ & \dots  & $23.5$ & \dots \\
        Decam $g$             & $0.473$ & $24.6$  & \dots & $24.6$ \\
        Decam $r$             & $0.642$ & $24.3$  & \dots & $24.3$ \\
        Decam $i$             & $0.784$ & $23.7$  & \dots   & $23.7$ \\
        Decam $z$             & $0.926$ & $22.9$  & \dots   & $22.9$ \\
        VIS \IE               & $0.715$ & $24.7$  & $24.7$ & $24.7$ \\
        NISP \YE              & $1.085$ & $23.1$  & $23.1$ & $23.1$ \\ 
        NISP \JE              & $1.375$ & $23.2$  & $23.2$ & $23.2$ \\
        NISP \HE              & $1.773$ & $23.2$  & $23.2$ & $23.2$ \\
        IRAC [IRAC1] & $3.550$ & $24.0$  & $24.0$ & $23.1$ \\
        IRAC [IRAC2] & $4.493$ & $23.9$  & $23.9$ & $23.0$ \\
        IRAC [IRAC3] & $5.696$ & $21.2$ & $20.0$  & \dots \\
        IRAC [IRAC4] & $7.799$ & $19.9$ & $21.1$  & \dots \\
        \hline

    \end{tabular}
\end{table}

To ensure the robustness of our sample, before applying our colour selection, we implemented a series of cuts:
\begin{enumerate}
    \item \texttt{SPURIOUS\_FLAG} $=0$;
    \item \texttt{DET\_QUALITY\_FLAG} $< 4$;
    \item \texttt{MUMAX\_MINUS\_MAG} $> -2.6$;
    \item $23.9 - 2.5 \, \logten$(\texttt{FLUX\_H\_TOTAL}) $< 24.5$;   
    \item \texttt{flag\_H} $= 0$.
\end{enumerate}

The first cut ensures the removal of all the objects that have been labelled as spurious in the official catalogue. Similarly, the second flag requires that the photometry is good, and the fourth one requires a magnitude below $24.5$ in the \HE band. The decision to apply a magnitude cut in the \HE band is due to the fact that this is one of the two bands used in our colour selection (see next paragraph). This is a conservative choice, in order to deal only with the brightest and most massive sources. The \texttt{MUMAX\_MINUS\_MAG} quantity, instead, represents a sort of estimate of the compactness of the sources, and this cut should remove all the stars present in the catalogue. The last condition ensures  that the \HE band is not affected by spurious detections or artefacts. Again, we decided to include it given that our selection mainly relies on this band. We show in Table \ref{tab:number_counts} the initial number of objects and the resulting number after applying these cuts. 

\begin{table}
    \centering
    \caption{Total number of objects in the merged \Euclid $+$ \textit{Spitzer} catalogue and the remaining number of objects after each cleaning step, shown for the three fields. The clean sample is the result after applying the cuts described in Sect. \ref{sc:methods}. The HIERO sample is obtained by applying the colour criterion $\HE - \mathrm{IRAC2} > 2.25$ in the clean sample. Lastly, the final HIERO sample is the result of our visual check (see Sect. \ref{visual_check}).}
    \label{tab:number_counts}
    \scalebox{0.9}{%
    \begin{tabular}{lccc}
    \hline
    \hline
    \noalign{\vskip 2pt}
         Field&  Fornax&  North& South\\
    \noalign{\vskip 2pt}
    \hline
    \noalign{\vskip 2pt}
         Original sample& $5\,328\,489$ & $11\,378\,352$ & $13\,060\,965$ \\
         
         Clean sample& $2\,837\,465$ & $5\,365\,713$ & $6\,534\,328$\\
         
         HIERO sample& $5\,263$ & $8\,258$ & $13\,385$ \\
         
         Final HIERO sample& $920$ & $1\,051$ & $1\,899$\\
         \hline
    \end{tabular}}
\end{table}

\begin{figure}
    \centering
    \includegraphics[width=1\linewidth]{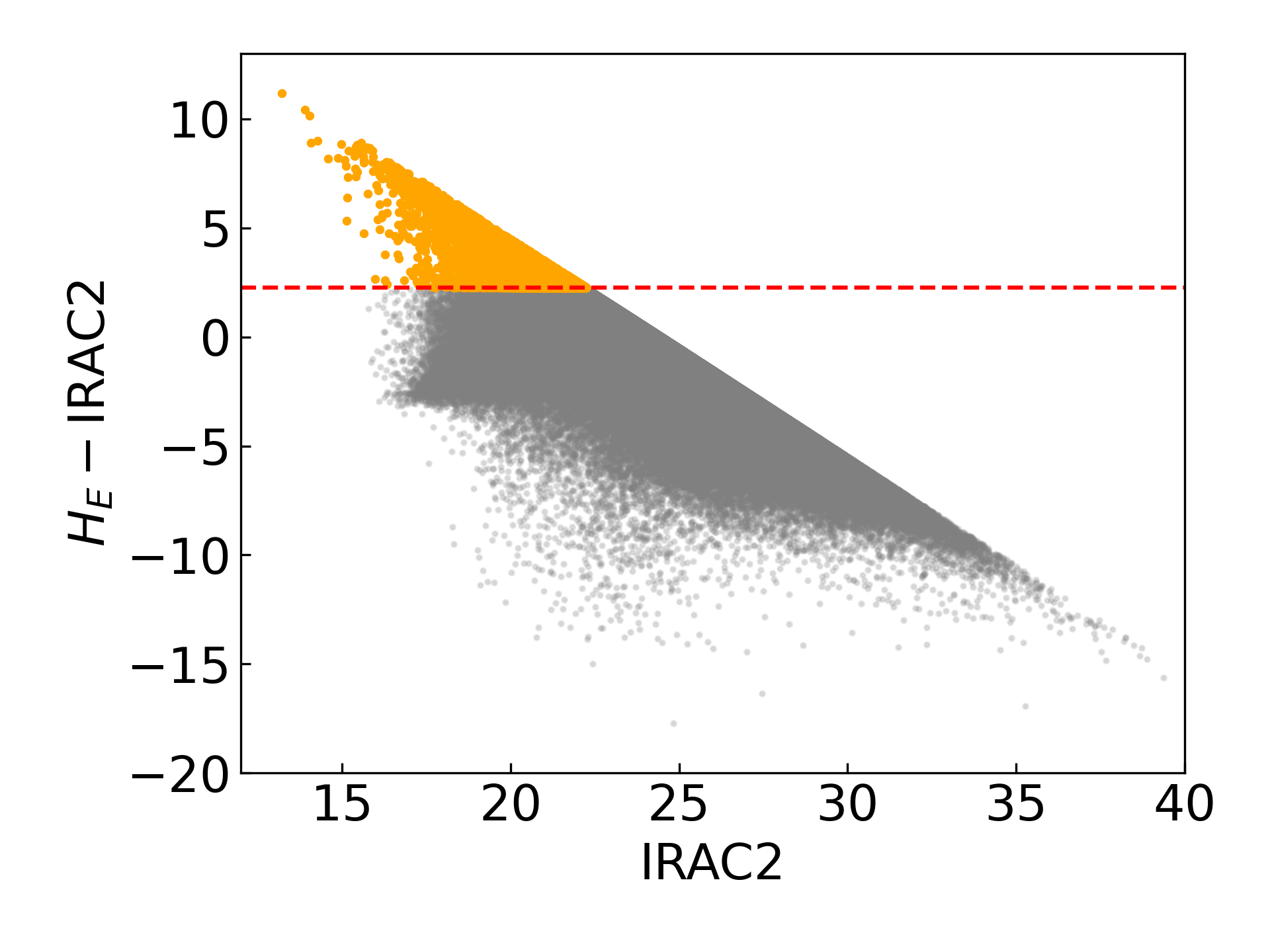}
    \caption{Colour-magnitude plot of the clean sample. All the object falling above the red line, representing $\HE - \mathrm{IRAC2} > 2.25$, i.e., the HIERO colour selection \citep{wang2016infrared}, end up being in our HIERO sample. The diagonal cut is due to the magnitude limit in the \HE band.}
    \label{fig:color-selection}
\end{figure}

To this clean catalogue, we applied the colour selection that identifies HIERO objects, as defined by \cite{wang2016infrared}, and originally introduced by \cite{Caputi12}: $\HE - \mathrm{IRAC2} > 2.25$. This colour selection is optimised to identify galaxies with $A_V \gtrsim 2$ mag and $\logten(M_* /M_\odot) \approx 10$ at $z \gtrsim 3$ \citep{gottumukkala2024unveiling}. Figure \ref{fig:color-selection} shows the entire clean parent sample, with the red line marking the $\HE - \mathrm{IRAC2} = 2.25$ limit to our selection. All the objects falling above this line respect the HIERO definition. Out of the $14\,737\,506$ objects in the clean sample, $26\,906$ candidates respect the colour criterion imposed and end up in our HIEROs sample.

%Given the worse spatial resolution by \textit{Spitzer}, which has a PSF $\sim 1$\arcsecond, some sources can have a contaminated IRAC photometry, due to the blending with close companions, invalidating the photometry retrieved. To remove this problem, we checked how many of these \HE-undetected sources have a detected source within a radius of $1$\arcsecond. Table \ref{tab:cuts} reports the number of objects after each removing step. 

\subsection{Visual investigation of the candidate HIEROs} \label{visual_check}
%Given the worse spatial resolution by \textit{Spitzer}, which has a PSF $\sim 1$\arcsecond, some sources can have a contaminated IRAC photometry, due to the blending with close companions, invalidating the photometry retrieved. To remove this problem, we checked how many of these \HE-undetected sources have a detected source within a radius of $1$\arcsecond. Table \ref{tab:cuts} reports the number of objects after each removing step. --> Non abbiamoo più H-undetected! Abbiamo messo il cut in H.
\begin{figure}
    \centering
    \includegraphics[width=1\linewidth, trim={0 0 600 60},clip, keepaspectratio]{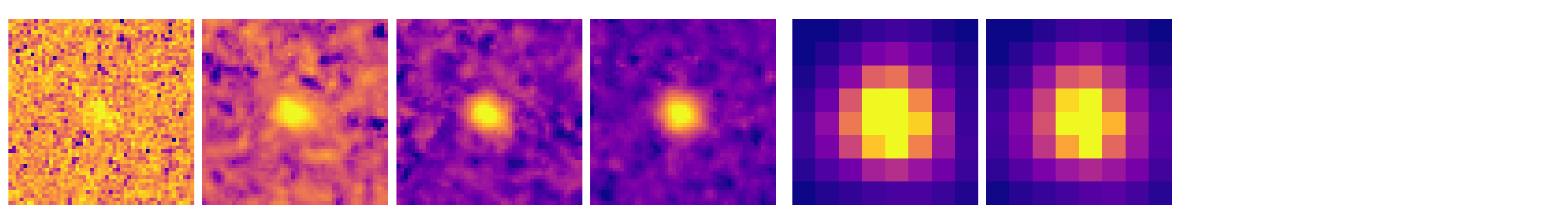}
    \includegraphics[width=1\linewidth, trim={0 0 600 60},clip, keepaspectratio]{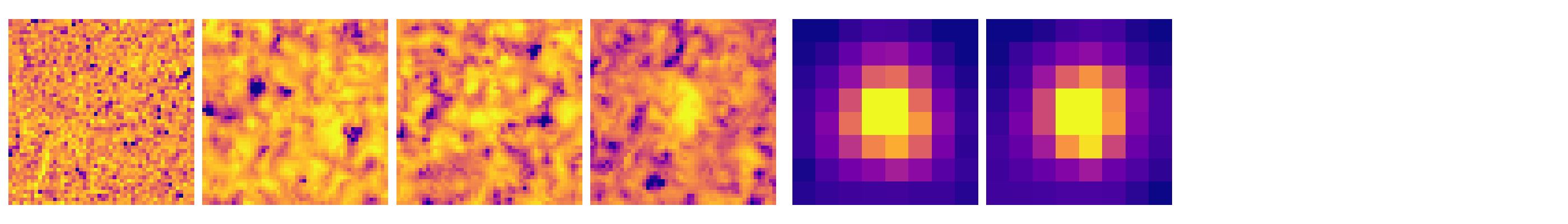}
    \caption{Examples of two HIEROs that have passed our visual check and that are kept in our final catalogue.
    From left to right: \IE, \YE, \JE, \HE, IRAC1, and IRAC2. Each cutout has a size of $\ang{;;5}\,\times\,\ang{;;5}$.}
    \label{fig:cutout-good}
\end{figure}

\begin{figure}
    \centering
    \includegraphics[width=1\linewidth, trim={0 0 600 60},clip, keepaspectratio]{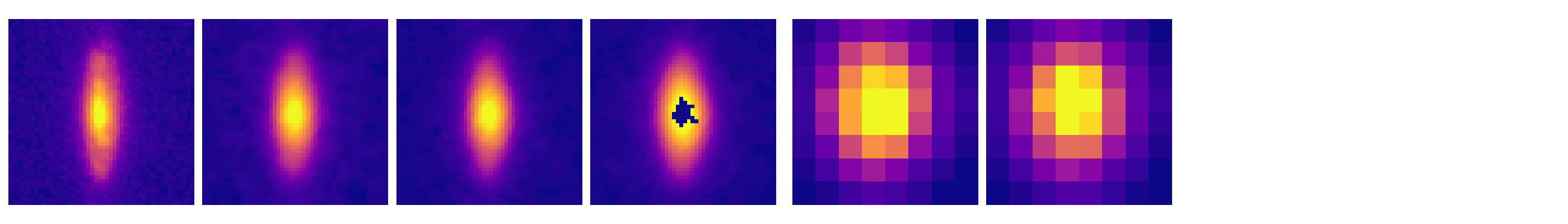}
    \includegraphics[width=1\linewidth, trim={0 0 600 60},clip, keepaspectratio]{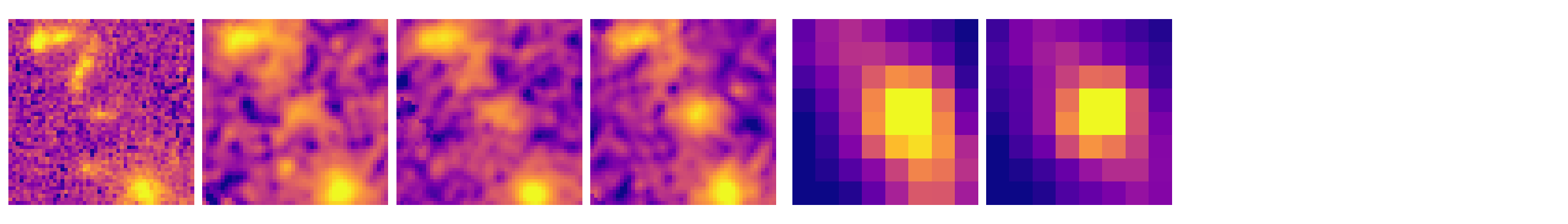}
    \includegraphics[width=1\linewidth, trim={0 0 600 60},clip, keepaspectratio]{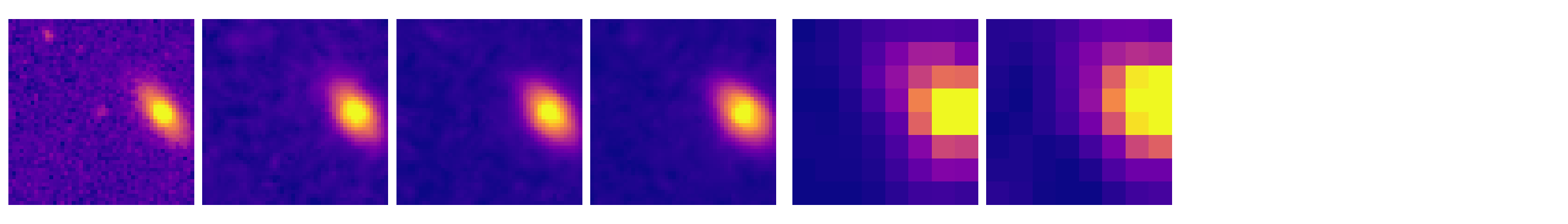}
    \caption{Examples of three HIEROs that did not pass our visual check and that are discarded from our final catalogue.
    From left to right: \IE, \YE, \JE, \HE, IRAC1, and IRAC2. Each cutout has a size of $\ang{;;5}\,\times\,\ang{;;5}$.}
    \label{fig:cutout-bad}
\end{figure}

To maximise the reliability of the sources in our sample, we have performed a visual check of all the HIERO selected according to the colour criterion presented in Fig. \ref{fig:color-selection}.
This is mandatory to account for various issues that could affect these data and in order to provide the most conservative sample for the statistical purposes of this paper.
We prioritize the purity at the expenses of the completeness. As reported in Sect. \ref{HIERO selection}, for a total of $26\,906$ objects we created a set of postage stamps, including the four \Euclid bands and the first two IRAC channels with a size of $\ang{;;5}\,\times\,\ang{;;5}$.

Some examples of good, isolated objects are reported in Fig.~\ref{fig:cutout-good}. We also highlight how the applied colour selection naturally includes dropout objects (bottom panel of Fig.~\ref{fig:cutout-good}). The difference in the point-spread functions (PSFs) of \Euclid and of \textit{Spitzer} (with full-width half maxima, FWHM, of the order of $\ang{;;1.5}$ in the first two IRAC channels) is  evident from the images. This immediately leads to a 
%brings to 
large uncertainty in the physical association of objects detected with the two instruments.
In particular, blending of \Euclid sources in the IRAC images is a major issue. Figure \ref{fig:cutout-bad} shows some such cases (middle panel),
together with other examples of HIERO candidates that we decided to discard. For example, automatic masking of pixels from the \Euclid pipeline can bias the measured fluxes in some bands, leading to artificially red galaxies (top panel). This does not imply that the object is not a valid candidate; rather, the available \Euclid photometry is insufficient to accurately recover its true colours. An alternative approach would have been to mask the same pixels across all bands. However, given our goal of constructing a conservative and robust sample, we opted to discard such objects instead.
Furthermore, the adopted IRAC photometry (see Sect. \ref{IRAC}), extracted at the position of the \Euclid sources, leads to cases where the corresponding IRAC flux is emitted by a different \textit{Spitzer} object (bottom panel in Fig. \ref{fig:cutout-bad}, where a faint VIS detection is visible at the centre of the cutout). We further discarded sources falling at the edges of the maps or dominated by any other evident artefacts.

To summarise, we determined whether to discard an object based on the following criteria:
\begin{enumerate}
\item the presence of bad pixels invalidating the photometry;
\item the presence of other \Euclid sources within the IRAC FWHM;
\item flux in the IRAC bands contaminated by nearby sources.
\end{enumerate}

Our conservative approach leads to a significant reduction in the number of HIERO sources retained in this study, with only $15$\% surviving the selection process. The exact values for each deep field are provided in Table \ref{tab:number_counts}, resulting in a final sample of $3\,870$ sources. While we acknowledge that discarding such a large number of objects is not ideal, the primary goal of this first work is to ensure a highly conservative selection. Consequently, we have not used this sample for statistical analyses, such as computing the stellar mass function, since completeness cannot be reliably reproduced. These analyses will be conducted once deblended IRAC photometry becomes available for all three EDFs.

\subsection{Physical parameters retrieval} \label{SED_fitting}
In this paper, we exploited the Python code \texttt{CIGALE} \citep{cigale}, given its fast response. We considered detections only the fluxes presenting S/N $>3$, for all the others we set the flux to $0$ and the error as $3$ times the observed depth reported in Table \ref{tab:filters}. The set up is reported in Table \ref{tab:parameters}.

\begin{table}
    \centering
     \caption{Input models and main parameters for the \texttt{CIGALE} code. The models used are: a delayed star-formation history with optional exponential burst; \cite{bruzual2003stellar} simple stellar population model; a continuum and line nebular emission model; a modified \cite{calzetti2000dust} dust attenuation law; and a redshifting model that also includes the intergalactic medium from \cite{meiksin2006colour}.}
     \label{tab:parameters}
   \begin{tabular}{lr}
    \hline
    \hline
    \noalign{\vskip 2pt}
    \multicolumn{2}{c}{\texttt{sfh\_delayed}} \\
    $\tau$ (main) [Myr] & $200, 300, 500, 700$ \\
    & $1000, 1500, 2000$\\
    Age (main) [Myr]& $200, 300, 500, 700$ \\
    & $1000, 1500, 5000$ \\
    \hline
        \noalign{\vskip 1pt}
    \multicolumn{2}{c}{\texttt{bc03}} \\
      IMF   & Chabrier\\
       Metallicity & $0.008, 0.02$\\
       \hline
               \noalign{\vskip 1pt}
       \multicolumn{2}{c}{\texttt{nebular}} \\
         logU& $-4.0, -3.0, -2.0, -1.0$\\
         Emission & True \\
       \hline
               \noalign{\vskip 1pt}
       \multicolumn{2}{c}{\texttt{dustatt\_modified\_starburst}} \\
        $E_{BV}$ lines [mag]& $0, 0.1, \dots, 4.4, 4.5$\\
        $E_{BV}$ factor& $0.44$\\
        $R_V$ & $3.1$ \\
       \hline
               \noalign{\vskip 1pt}
        \multicolumn{2}{c}{\texttt{redshifting}} \\
        Redshift & $0, 0.1, \dots, 14.9, 15$\\
        \hline
    \end{tabular}

    \end{table}

We mimicked the set up used in the analogous work with the ERO data (Girardi et al. in prep.). The $A_V$ value is up to $6$, since we expect very dusty sources; while the redshift is free to go up to $15$. This is because we expect both low and high-$z$ contaminants. With the wide nebular parameter range we ensure that we are not overestimating the mass due to mistaking the emission lines as the continuum, given the absence of data at longer wavelengths that could mitigate this problem. This could lead to an overestimation by up to a factor of $10$ \citep{bisigello2019statistical, papovich2023ceers}.

\section{Results and Discussion} \label{sc:Results}
Given the significant uncertainties and degeneracies associated with photometric redshift estimation, we focus on sources with at least three detections. When only one or two photometric points are available, even when considering upper limits, the fits cannot be considered reliable. Despite this restriction, the large sample size still allows for meaningful analysis. Prioritising robustness over completeness, we exclude these sources from further discussion. The resulting final sample consists of $2\,994$ galaxies.

However, despite removing sources with the fewest photometric detections, we stress that a fraction of our sample still relies on SED fits constrained by only three or four detections with S/N $> 3$. Given these limitations, we invite the reader to interpret the derived photometric redshifts and galaxy properties with caution. This uncertainty is reflected in the non-negligible errors, reported in each plot. Future observations of the EDFs will be crucial for these objects, because they will benefit from the increased depth of upcoming surveys.

\begin{figure*}
    \centering
    \includegraphics[width=0.49\linewidth]{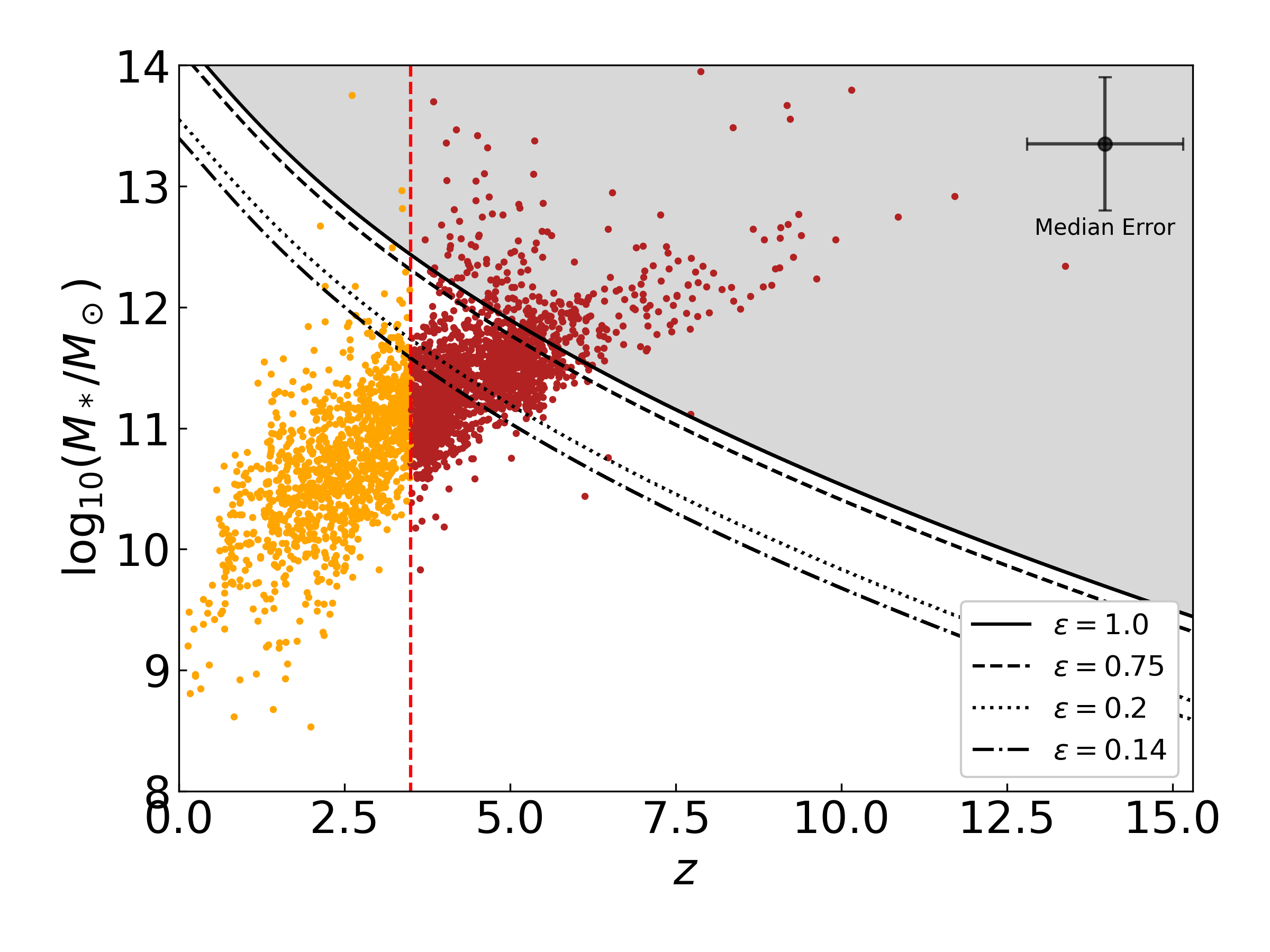}
    \includegraphics[width=0.49\linewidth]{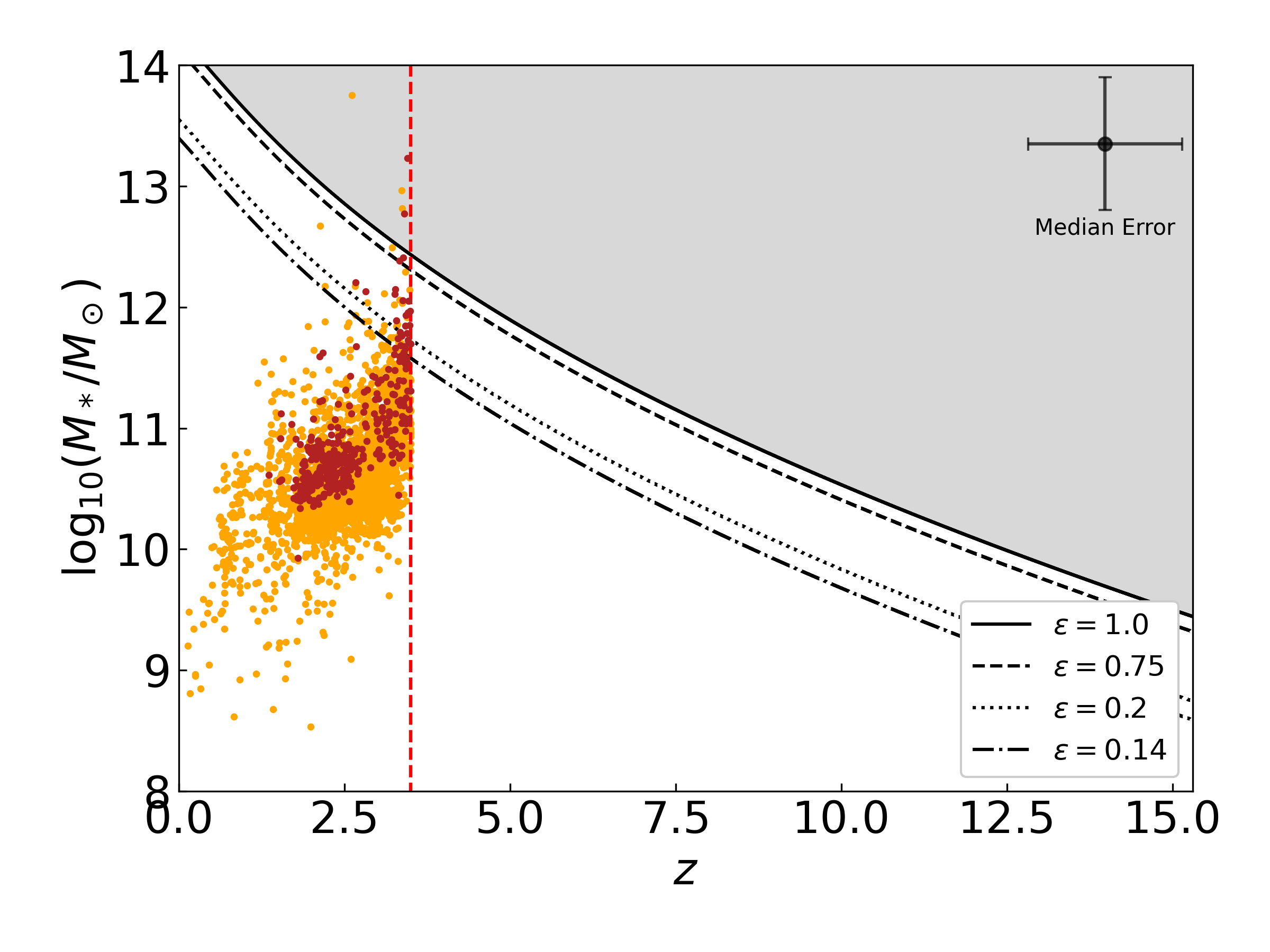}
    \caption{Stellar mass versus redshift distribution of the HIERO sample. The \textit{y}-axis is on a logarithmic scale. The vertical red dotted line corresponds to $z = 3.5$, above which we don't trust the high redshift solutions found by the fit. These high-$z$ sources are displayed as red dots. The grey-shaded area represents the region forbidden by the $\Lambda$CDM model. The black lines show the limit for different values of $\epsilon$. The median error for both the quantities is reported in the top right corner of the plot. The left panel show the results from the \texttt{CIGALE} run described in Sect. \ref{SED_fitting}, while the right panel uses the results from the low-$z$ run for the high-$z$ objects, which are still displayed as red dots.    
    }
    \label{fig:M-vs-z}
\end{figure*}

We first present the $M_*$ versus $z$ distribution in Fig. \ref{fig:M-vs-z}. The left panel displays all data points for sources with three or more detections, using values derived from \texttt{CIGALE}. The grey-shaded region represents the prohibited area according to the $\Lambda$CDM cosmological model, where baryonic conversion efficiency ($\epsilon$) exceeds $100\%$. Approximately $9.7\%$ of the sample falls within this region. The vertical red dotted line marks $z=3.5$, with sources to the right of this line represented by red dots; we refer to these as high-$z$ sources from this point onward. Given the limited photometric information and the relatively shallow depth of this data release (as reported in Table \ref{tab:filters}), we do not place full confidence in the SED-fitting results for these objects.

We also take into account the findings of \cite{forrest2024magaz3ne}, whose empirical study suggests that SED-fitting estimates yielding $M_* > 10^{11.7}\,M_\odot$ at $3<z_{\mathrm{phot}}<4$ are not supported by spectroscopic observations. Specifically, they selected a sample of red objects with such estimated properties and obtained spectroscopic data. Upon comparison, none of the objects satisfied $z_{\mathrm{spec}} - z_{\mathrm{phot}} < 0.5$, underscoring the difficulty in accurately characterising such sources. Their results, based on very high S/N observations and a broader set of observed bands, rely on more robust photometry. Given this distinction, we opted not to impose a stellar mass cut but instead to focus solely on redshift considerations.

\begin{figure}
    \centering
    \includegraphics[width=1\linewidth]{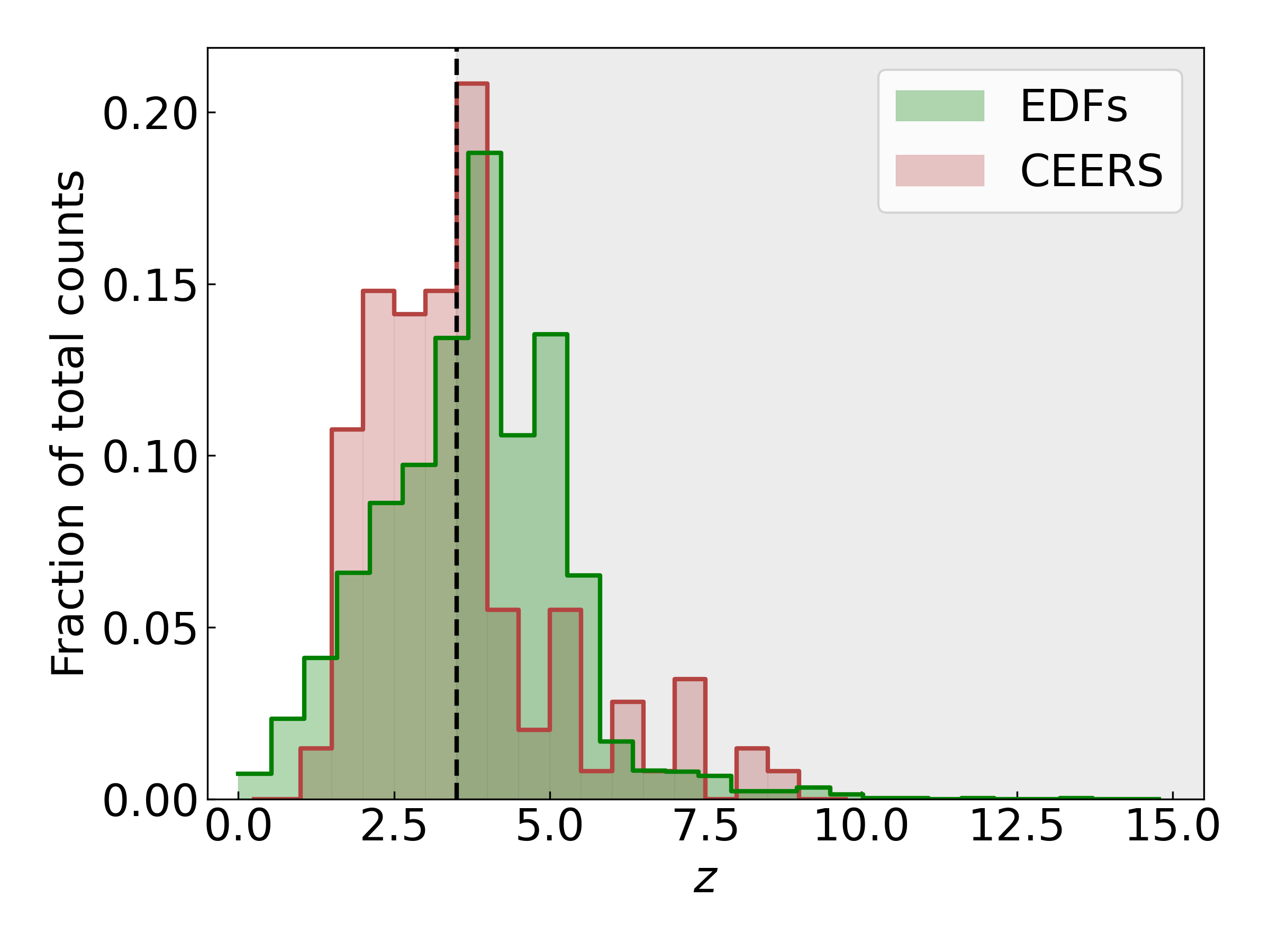}
    \caption{Redshift distributions for the total HIERO sample in the EDFs in green, compared to the distribution found by \cite{gottumukkala2024unveiling} in the CEERS field applying a similar selection, displayed in red. The vertical line corresponds to $z = 3.5$. }
    \label{fig:z-distribution}
\end{figure}

This decision is further supported by an analysis of the total redshift distribution of the EDFs, normalised by total counts and shown in Fig. \ref{fig:z-distribution}. We find that our number counts at $z>3.5$ exceed those reported by \cite{gottumukkala2024unveiling} in the CEERS field. The latter study utilises JWST observations covering a significantly smaller area (about $80\,\mathrm{arcmin}^2$) but reaching considerably greater depth than the current Q$1$ data release. This discrepancy raises concerns about the reliability of our photometric redshift estimates for the high-$z$ sub-sample. To address this, we performed a second \texttt{CIGALE} run, constraining the redshift to $z<3.5$. The right panel of Fig. \ref{fig:M-vs-z} presents the same distribution, but incorporates the results from this second run for high-$z$ sources, which are still shown as red dots to illustrate their redistribution. We observe that these empirical constraints align well with the theoretical limit, reducing the fraction of points in the prohibited area to approximately $0.17\%$. Given this refined distribution, we do not blindly trust the SED-fitting tool’s results, but instead integrate both empirical and theoretical insights. 

Preferring a lower-$z$ solution for this sub-sample implies prioritising a low-$z$ interpretation over a high-$z$ one. In fact, in the preliminary work of Girardi et al. (in prep.), we found that nearly all HIEROs exhibit a bimodal or multi-peak probability distribution function for redshift. This is due to the degeneracy between a high-$z$, less dusty solution and a lower-$z$, more dusty one. Statistically, some objects are likely to be genuine high-$z$ sources; however, given our inability to discern which cases reflect true high-$z$ objects versus those with overestimated photometric redshifts, we adopt a conservative approach. The presence of high-$z$ sources could be further tested in a future work with longer wavelength data, such as submillimeter observations from Herschel/SPIRE. Based on all these considerations, we proceed by presenting and discussing results derived from the distribution shown in the right panel of Fig. \ref{fig:M-vs-z}.

\begin{figure}
    \centering
    \includegraphics[width=1\linewidth]{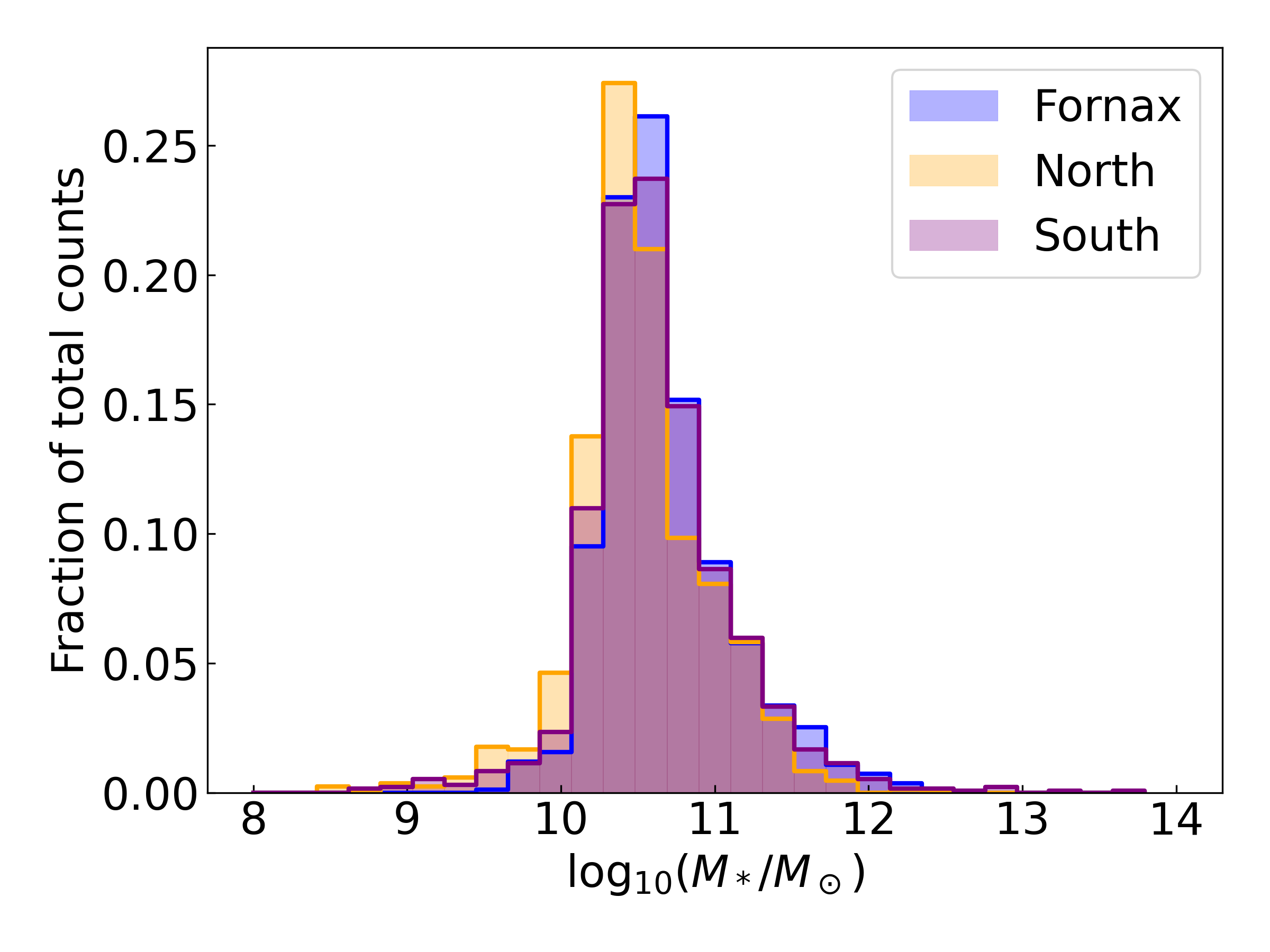}
    \caption{Stellar mass distributions for the EDF-F in blue, EDF-N in orange, and EDF-S in purple. The distributions have the counts normalised to the total number of objects in each field.    
    }
    \label{fig:M}
\end{figure}

\begin{figure}
    \centering
    \includegraphics[width=1\linewidth]{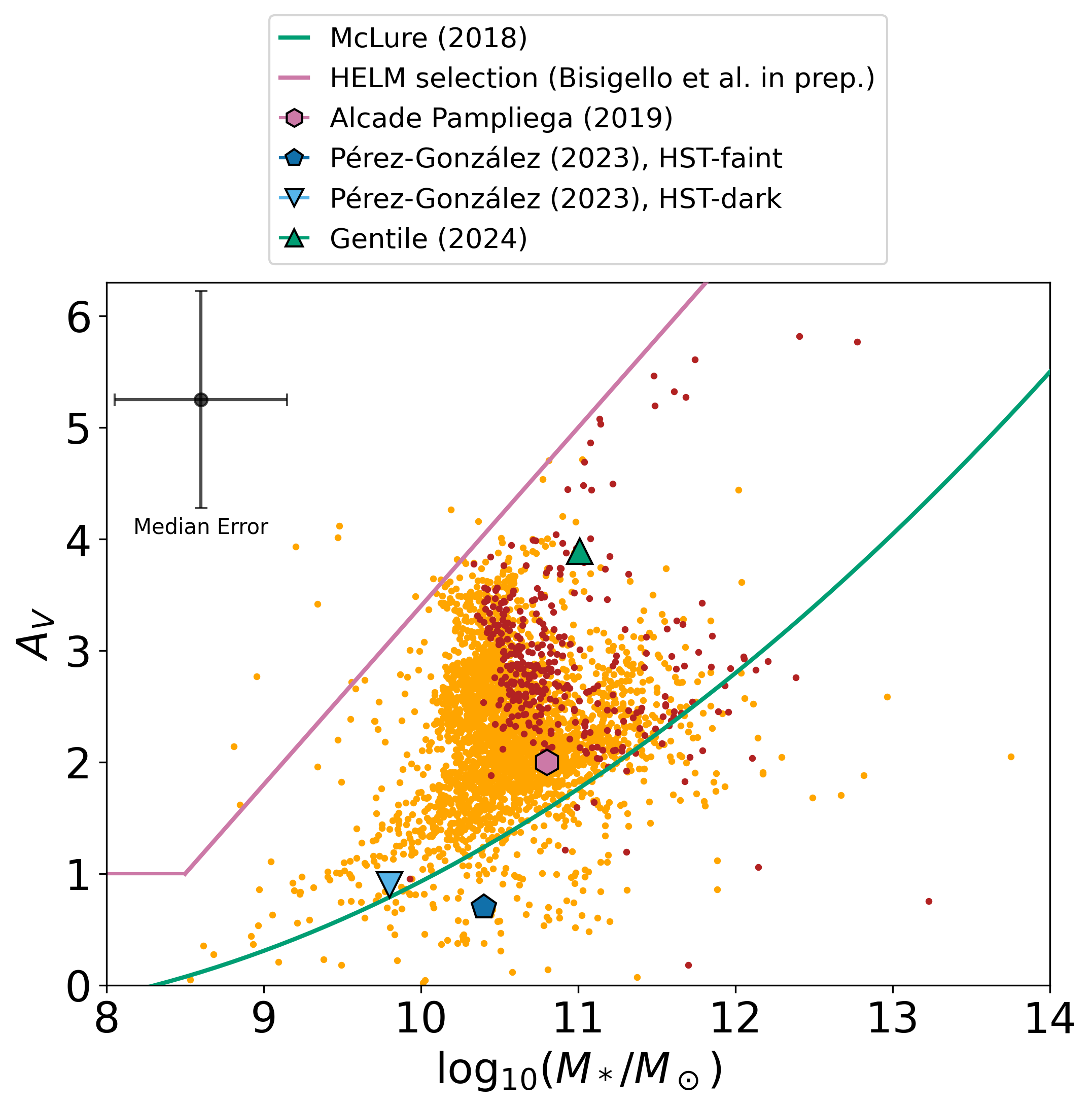}
    \caption{Dust attenuation versus stellar mass of the HIERO sample. The \textit{x}-axis is on a logarithmic scale. The dark teal solid line reports the relation of \cite{mclure2018dust}, while the purple line delimits the area that identifies the so-called HELM galaxies \citep{bisigello2024spectroscopic}. The median error for both the quantities is reported in the top left corner of the plot. Different symbols report values from previous studies, as indicated in the legend.     
    }
    \label{fig:Av}
\end{figure}

%The three distributions are quite in agreement, however they present some trends that differ from one to another. EDF-F distribution has a peak at $z \sim 4$ that is roughly $2 \times$ higher than EDF-S and $\sim 5 \times$ EDF-N peak. Even considering the different initial number of objects in the three fields, EDF-S distribution seems to prefer solutions at lower $z$ with respect to the two other fields. 

%When comparing this distribution to the one obtained by plotting the photometric redshift of the entire final sample, we notice that the extremes of the distribution ($z\sim0$ and $z>6$) disappears. In particular, the EDF-S peak at $z\sim0$ in the whole sample is $10$ times larger than the one visible in the top panel of Fig. \ref{fig:z-distribution_sub-sample}. Similarly, nothing appears at $z > 6$, while in the whole sample case there were some objects. This shows that the objects belonging to the sub-sample where the \texttt{nnpz} procedure did not work well are at the extremes of our redshift parameter space. 

Figure \ref{fig:M} shows the distribution of the stellar masses. The counts are proportional to the different numbers of candidates in the different fields. We find a mean value of $\langle M_* \rangle = 10^{10.6}\,M_\odot$, a value that confirms the expected massive nature of these sources.

Studying this type of source, we are interested in looking at the dust attenuation values, since we expect them to be very dusty. We find a mean value of $\langle A_V \rangle = 2.3$, confirming the expectation. Looking at the $A_V$ versus $M_*$ distribution, shown in Fig. \ref{fig:Av}, we see that our points are spread across quite a  broad range, covering  $A_V$ from about $0$ up to $6$. For reference, the expected relation for normal galaxies in \cite{mclure2018dust} is also shown as a dark teal solid line. Our sample roughly follows this relation, revealing, however, a possible population of extremely obscured objects at $M_*>10^{10}\,M_{\odot}$. This proves the power of \Euclid in providing statistical samples of candidates of such rare objects for subsequent follow up observations using other facilities, in order to confirm their redshifts and nature. For comparison, we include in this plot recent results from the literature of samples selected with similar criteria \citep{pampliega2019optically, perez2023ceers, gentile2024dark}, including much fainter JWST dark objects.

%However, given the wider sky area, we also find very massive objects in the EDFs, i.e. $<M_*> \sim 10^{11}\,M_\odot$ in all  three fields, reaching the likely unphysical values of $10^{13}\,M_\odot$, both in EDF-F and EDF-S. To better connect the different parameters, Fig. \ref{fig:physical_param} (top panel) shows the distribution of the stellar masses versus redshift. For comparison, the stellar mass predicted for three star-forming galaxies with different ages between the redshift epoch and the peak of the burst ($50$, $100$, and $200$, respectively) are reported. This comparison allows us to notice that our sample mainly lies above these solid lines, therefore confirming that we are observing a massive sample. 

%\textcolor{purple}{Interestingly, a small fraction of candidates (around $4.5\%$) sit  inside the purple solid line that delimits the highly extinguished low-mass (HELM) objects. These galaxies are defined and some are spectroscopically confirmed in the work of \cite{bisigello2024spectroscopic}. }

%Our values cover all the reference from different works studying objects with similar selection criteria and expected physical parameters, i.e. \cite{pampliega2019optically, perez2023ceers, gentile2024dark}, showing how our sample does not present characteristics confined in a narrow physical properties range. This underscores the importance of delving deeper to better understand this population, which still requires comprehensive characterization. For this reason, spectroscopic observations are indeed essential to draw robust conclusions on their nature. 

\section{Conclusions}

In this paper, we have exploited the Q$1$ \Euclid data release to characterise HIERO galaxies, a dusty and massive population exhibiting extremely red colours. Starting from the official \Euclid photometric catalogue, we applied a series of selection cuts (see Sect. \ref{HIERO selection}) to isolate candidates that meet the HIEROs colour criteria. This process resulted in a clean sample of $26\,906$ sources.

To ensure the robustness of our sample, we performed a rigorous visual inspection, retaining only candidates with reliable photometry (i.e., free from defects, masked regions, or blending/contamination issues in the IRAC bands; see Sect. \ref{visual_check}). This refinement led to a final sample of $3\,870$ sources.

We used an SED-fitting code to derive the physical properties of these objects (see Sect. \ref{SED_fitting}). To enhance the robustness of our sample, we included only sources with at least three photometric detections (S/N $>3$). However, the results should be interpreted with caution, as in some cases, the SED fitting relies on only three or four photometric points. Nevertheless, we believe these findings are valuable, both for anticipating the potential of future \Euclid data releases and as foundation for further investigations.

The wide area of the EDFs allowed us to obtain meaningful statistics, even after applying all the selection cuts to define the final sample. This is particularly crucial for these rare objects, which would otherwise be challenging to characterise. The next step is to leverage future data releases to accurately determine the HIERO contribution to the stellar mass function across different epochs, a task that will be addressed in a forthcoming paper.

This pilot analysis highlights the need for further studies to fully understand this population. Our results show that these objects span a broad range of parameters, leaving their nature still uncertain. Obtaining spectroscopic data will be crucial to better constrain their properties. \Euclid's slitless spectroscopy will help disentangle the degeneracy between redshift and dust attenuation, determining whether the identified high-$z$ contaminants are genuine or instead represent more dust-obscured sources at lower redshifts.

Overall, \Euclid has demonstrated its potential as a pivotal instrument for studying this massive and dusty population. Even more exciting results are expected with the DR$1$ \Euclid data release, which will cover approximately $1900\,\mathrm{deg}^2$ of sky.

%
% Add the acknowledgement using the achnowledgements environment.
% Do not use \acknowledgement{....} as this affects the formatting
% of the references.
%

\begin{acknowledgements}
%The authors acknowledge support from the ELSA project.``ELSA: Euclid Legacy Science Advanced analysis tools" (Grant Agreement no.~101135203) is funded by the European Union. Views and opinions expressed are however those of the author(s) only and do not necessarily reflect those of the European Union or Innovate UK. Neither the European Union nor the granting authority can be held responsible for them. UK participation is funded through the UK Horizon guarantee scheme under Innovate UK grant 10093177.
The research activities described in this paper were carried out with contribution of the Next Generation EU funds within the National Recovery and Resilience Plan (PNRR), Mission 4--Education and Research, Component 2--From Research to Business (M4C2), Investment Line 3.1--Strengthening and creation of Research Infrastructures, Project IR0000034--``STILES--Strengthening the Italian Leadership in ELT and SKA”.
%\AckERO  
\AckQone
\AckEC  
\AckDatalabs
\end{acknowledgements}

%
% Here comes the reference list, generated via bibtex from
% your bibfile my.bib and Euclid.bib. Please make sure that
% the same paper is not referenced twice, one from your my.bib
% file, and once from Euclid.bib.
%

\bibliography{Euclid, Q1, my} % add my.bib, containing your bibentry file 

%
% Now you can add appendices.
% In this example, the appendices are in one column mode.
% If that is not requires, comment out \onecolumn
% Note that appendices in A\&A come {\it after\/} the references.

%\begin{appendix}
%  \onecolumn %If you don't want single column for the Appendix, please
             %comment this out

%
%\end{appendix}

\label{LastPage}
\end{document}